\documentclass[12pt,epsf]{article}
\usepackage{amsmath, amssymb}
\usepackage{graphicx}
\begin{document}

\def\a{\alpha}
\def\b{\beta}
\def\c{\varepsilon}
\def\d{\delta}
\def\e{\epsilon}
\def\f{\phi}
\def\g{\gamma}
\def\h{\theta}
\def\k{\kappa}
\def\l{\lambda}
\def\m{\mu}
\def\n{\nu}
\def\p{\psi}
\def\q{\partial}
\def\r{\rho}
\def\s{\sigma}
\def\t{\tau}
\def\u{\upsilon}
\def\v{\varphi}
\def\w{\omega}
\def\x{\xi}
\def\y{\eta}
\def\z{\zeta}
\def\D{\Delta}
\def\G{\Gamma}
\def\H{\Theta}
\def\L{\Lambda}
\def\F{\Phi}
\def\P{\Psi}
\def\S{\Sigma}

\def\o{\over}
\def\beq{\begin{eqnarray}}
\def\eeq{\end{eqnarray}}
\newcommand{\gsim}{ \mathop{}_{\textstyle \sim}^{\textstyle >} }
\newcommand{\lsim}{ \mathop{}_{\textstyle \sim}^{\textstyle <} }
\newcommand{\vev}[1]{ \left\langle {#1} \right\rangle }
\newcommand{\bra}[1]{ \langle {#1} | }
\newcommand{\ket}[1]{ | {#1} \rangle }
\newcommand{\EV}{ {\rm eV} }
\newcommand{\KEV}{ {\rm keV} }
\newcommand{\MEV}{ {\rm MeV} }
\newcommand{\GEV}{ {\rm GeV} }
\newcommand{\TEV}{ {\rm TeV} }
\def\diag{\mathop{\rm diag}\nolimits}
\def\Spin{\mathop{\rm Spin}}
\def\SO{\mathop{\rm SO}}
\def\O{\mathop{\rm O}}
\def\SU{\mathop{\rm SU}}
\def\U{\mathop{\rm U}}
\def\Sp{\mathop{\rm Sp}}
\def\SL{\mathop{\rm SL}}
\def\tr{\mathop{\rm tr}}

\def\IJMP{Int.~J.~Mod.~Phys. }
\def\MPL{Mod.~Phys.~Lett. }
\def\NP{Nucl.~Phys. }
\def\PL{Phys.~Lett. }
\def\PR{Phys.~Rev. }
\def\PRL{Phys.~Rev.~Lett. }
\def\PTP{Prog.~Theor.~Phys. }
\def\ZP{Z.~Phys. }


\baselineskip 0.7cm
\begin{titlepage}
\begin{flushright}
IPMU12-0091\\
ICRR-report-615-2012-4\\
\end{flushright}

\vskip 1.35cm
\begin{center}
{\large \bf 
Seesaw Mechanism with Occam's Razor
}
\vskip 1.2cm
Keisuke Harigaya$^1$, Masahiro Ibe$^{1,2}$ and Tsutomu T. Yanagida$^1$
\vskip 0.4cm
$^1${\it Kavli IPMU, TODIAS, University of Tokyo, Kashiwa 277-8583, Japan}\\
$^2${\it ICRR, University of Tokyo, Kashiwa 277-8582, Japan}
\vskip 1.5cm

\abstract{
We discuss the seesaw mechanism 
which includes the minimum number of parameters 
for successful leptogenesis and three neutrino oscillations 
in the spirit of Occam's razor.
We show that models with two right-handed neutrinos with two texture zeros
supported by Occam's razor 
cannot fit the observed neutrino parameters consistently
for the normal light neutrino mass hierarchy.
For the inverted light neutrino mass hierarchy, on the other hand,
we find that the models can fit  the observed neutrino parameters consistently.
Besides, we show that the model predicts the maximal Dirac $CP$-phase of the neutrino mixing matrix 
in the measurable range in the foreseeable future for the inverted neutrino  mass hierarchy.
We also show that the predicted effective Majorana neutrino mass responsible for the neutrinoless
double beta decay is around $50\,$meV which is also within reach
of future experiments. 
}
\end{center}
\end{titlepage}

\setcounter{page}{2}
\section{Introduction}
The seesaw mechanism\,\cite{seesaw} is the most  fruitful explanation of
the light neutrino masses and  mixings,
which connects the tiny neutrino masses to very heavy right-handed neutrino masses.
In fact, the seesaw mechanism is promised to accommodate the 
experimental results of the neutrino oscillation experiments since the mechanism 
involves more parameters than can be determined by the neutrino 
masses and mixings.
The seesaw mechanism is also highly motivated
since it provides the origin of the baryon asymmetry of the universe
by leptogenesis\,\cite{leptogenesis}.

Recently, the third, but the last neutrino mixing angle $\theta_{13}$ has been measured 
at Daya~Bay\,\cite{An:2012eh} and RENO\,\cite{Ahn:2012nd} 
reactor neutrino oscillation experiments with very high accuracy. 
Thus now, we know three neutrino mixing angles, $\theta_{12}$, $\theta_{13}$ and $\theta_{23}$ 
and two squared-mass differences of neutrinos, ${\mit\D} m_{21}^2$ and ${\mit \D} m_{31}^2$. 
The remaining very important parameters in the neutrino sector are 
the effective Majorana neutrino mass, $m_{ee}$, which is responsible 
for the neutrinoless double beta decay,  and the Dirac $CP$-violating phase $\delta$.

Unfortunately, the seesaw mechanism is  impossible to predict those 
parameters because of  the lack of 
convincing theory of the origin of the Yukawa interaction matrices. 
Without any clues, the number of free parameters in the neutrino mass matrix 
is larger than the one of the observable parameters\,\cite{Branco:2002ie}.
In order to have predictabilities on the neutrino mass matrix,
one may assume some (discrete) symmetries among generations
(for a review see \cite{King:2003jb} and references therein). 
Unfortunately, however, no convincing and successful symmetries have been found
in light of the precise observations of $\theta_{13}$.

In this paper, alternatively, we discuss the seesaw mechanism in the sprit of Occam's razor.
That is, we allow the minimum number of parameters needed 
for successful leptogenesis and three neutrino oscillations\,\cite{Frampton:2002qc}.
Concretely, we consider the seesaw mechanism with only two right-handed neutrinos. 
We further restrict models so that the Yukawa coupling constants 
between the right-handed neutrinos,
the Standard Model leptons and the Higgs boson have two texture 
zeros\,\cite{Frampton:2002qc,Ibarra:2003up,Raidal:2002xf}. 
Notice that such restricted models have enough number of parameters for leptogenesis 
as well as for the three neutrino oscillations.
The first attempt of  imposing  zeros to the quark mass matrices
was made by Weinberg long time ago\,\cite{Weinberg}, 
which leads to a very successful result. 
There, the Cabibbo angle is predicted to be $\sqrt{m_d/m_s}\simeq 0.22$.
In our argument, we use the ideas of imposing zeros to the Yukawa coupling matrix
to make the seesaw mechanism as economical as possible,
and hence, predictable on the remaining two important parameters $\delta$ and $m_{ee}$.

As we will show, however, the  seesaw mechanism with the minimal number of parameters
achieved by two texture zeros cannot fit the five observed neutrino parameters consistently
for the normal light neutrino mass hierarchy.
For the inverted light neutrino mass hierarchy, on the other hand,
we find that the models can fit  the five observed neutrino parameters consistently.
Furthermore, we find that the model predicts the Dirac $CP$-phase, $\delta\simeq \pm \pi/2$,
which is in the measurable range in the foreseeable future.
Notice that the models with two texture zeros have only one $CP$-violating phase.
Thus, the $CP$-asymmetry for baryogenesis and the $CP$-violations 
in the neutrino oscillations are inevitably interrelated with each other.
We also show that the effective Majorana neutrino mass responsible for the neutrinoless
double beta decay is predicted to be around $50\,$meV which is also within reach
of future experiments.

The organization of the paper is as follows.
In section\,\ref{sec:parameters}, we briefly review the seesaw mechanism
and summarize the parametrization of the models.
In section\,\ref{sec:twozeros}, we derive the conditions for two texture zeros.
In section\,\ref{sec:normal}, we show  that the minimal seesaw mechanism 
cannot explain all the five neutrino parameters consistently for the normal light neutrino mass hierarchy.
In section\,\ref{sec:inverted}, we show that the models of the minimal seesaw
can consistently fit all the five observed neutrino parameters.
There, we also discuss the predictions on the Dirac $CP$-phase $\delta$
and the effective Majorana neutrino mass $m_{ee}$.
In section\,\ref{sec:leptogenesis}, we discuss numerical implications of the models on thermal leptogenesis.
The final section  is devoted to conclusions and discussions.

\section{Neutrino Parameters in Seesaw Mechanism}
\label{sec:parameters}
In the seesaw mechanism, we introduce $n_N$ 
Majorana right-handed neutrinos $N_i\,(i=1- n_N)$
which are singlets under the Standard Model gauge symmetries. 
The Lagrangian responsible for the seesaw mechanism is given by,
\begin{eqnarray}
\label{eq:lagrangian}
 {\cal L} = y_{\a\b}\ell_{L\a} \bar{e}_{R\b}h+ \lambda_{i \a} N_i \ell_{L\a} h - \frac{1}{2} M_{ij} N_i N_j \ ,
\end{eqnarray}
where $h$ denotes the Higgs doublets,  $\ell_{L\a}\, (\a = 1-3)$ the three generations of the
lepton doublets, and $\bar{e}_{R\a}\, (\a = 1-3) $ the three generations of the left-handed anti-leptons.
The coefficients $y$ and  $\lambda$ are the Yukawa coupling constants and $M$'s are the masses of 
the right-handed neutrinos.
Hereafter, we take  a basis where the masses of the right-handed neutrinos
and the Yukawa coupling constants $y_{\a\b}$  are diagonal, i.e. 
$M_{ij}= M_i \delta_{ij}$ and $y_{\a\b} = y_{\a}\delta_{\a\b}$.
Through the above Lagrangian, 
the tiny neutrino masses are generated by integrating out the heavy 
right-handed neutrinos,
\begin{eqnarray}
\label{eq:seesaw}
  (m_\n)_{\a\b}  = \sum_{i=1}^{n_N}\lambda^T_{\a i} M_i^{-1} \lambda_{i \b}\, v^2\ ,
\end{eqnarray}
where $v \simeq 174.1$\,GeV is the vacuum expectation value of the Higgs boson.

In applying Occam's razor to seesaw mechanism, let us first remind ourselves 
that at least two right-handed heavy neutrinos are necessary for the generation of the
baryon asymmetry of the universe\,\cite{Frampton:2002qc}.
Thus, we assume here only two right-handed neutrinos,  $n_N =2$, 
in the spirit of Occam's razor. 
It should be also noted that two right-handed neutrinos are 
also necessary to explain the two observed squared mass differences, ${\mit \D}m_{21}^2$ 
and ${\mit \D}m_{31}^2$.

In the models with two right-handed neutrinos, the generic Yukawa coupling constants 
form a $2\times 3$ complex valued matrix.
For later purpose, we define $2\times 2$ and $3\times 3$ diagonal
mass matrices of the right-handed neutrinos and the light neutrinos,
\begin{eqnarray}
 M_R = \diag(M_1, M_2)\ ,\qquad
 \bar{m}_\nu = \diag(m_1,m_2, m_3)\ ,
\end{eqnarray}
where we take all the mass parameters are real and positive without losing generality. 
We also arrange $M_1<M_2$ and $m_1 < m_2$.
The first important prediction of the models  with two right-handed neutrinos
is that  one of the light neutrino masses is vanishing, since the 
rank of the mass matrix $m_\nu$ in Eq.\,(\ref{eq:seesaw})
is $2$ for $n_N = 2$.
Thus, the model predicts $m_1 = 0$ for the normal neutrino mass hierarchy 
and $m_3 = 0$ for the inverted neutrino mass hierarchy.

The diagonalized neutrino mass matrix $\bar{m}_\nu$ 
is related to $m_\n$ in Eq.\,(\ref{eq:seesaw})
by the neutrino mixing matrix (MNS matrix)\,\cite{Maki:1962mu};
\begin{eqnarray}
 U_{MNS} &=&
\left(
\begin{array}{ccc}
U_{e1}  & U_{e2}  & U_{e3}   \\
U_{\m 1}  & U_{\m 2}  & U_{\m 3}   \\
U_{\t 1}  & U_{\t 2}  & U_{\t 3}   
\end{array}
\right)\ ,
\nonumber \\
&=&
\left(
\begin{array}{ccc}
c_{12}c_{13}  & s_{12} c_{13}   & s_{13} e^{-i \d}   \\
- s_{12}c_{23}-c_{12}s_{23}s_{13}e^{i\d}  & c_{12}c_{23}-s_{12}s_{23}s_{13}e^{i\d}   & s_{23}c_{13}   \\
 s_{12}s_{23}-c_{12}c_{23}s_{13}e^{i\d}  &  -c_{12}s_{23}-s_{12}c_{23}s_{13}e^{i\d}  &   c_{23}c_{13} 
\end{array}
\right)\nonumber\\
&&\times \diag(1,e^{i\a/2},1)\ ,\\
\bar{m}_\n &=& U_{MNS}^T\, \l^T\, M_R^{-1} \, \l \, U_{MNS}\, v^2\ .
\end{eqnarray}
Here, the sines and the cosines of the mixing angles $\theta_{ij}$ are abbreviated 
as $s_{ij} = \sin\theta_{ij}$ and $c_{ij} = \cos\theta_{ij}$.
In the mixing matrix, we have eliminated one of the Majorana phases in the neutrino mixing matrix 
since either $m_1$ or $m_3$ is vanishing for $n_N = 2$.

Now, let us compare the number of the parameters in the high and the low energy physics. 
The number of real parameters included in the light neutrino masses and the 
mixing matrix is seven; two neutrino masses, three mixing angles, and two phases.
This is less than the number of the parameters included in $\lambda$ which adds up to nine
after eliminating the three phases by rotating $\ell_{L}$ and $\bar{e}_R$.

To see how the two excessive parameters in $\l$ are hidden in the light neutrino mass matrix
in Eq.\,(\ref{eq:seesaw}), it is transparent to write a generic solution 
of Eq.\,(\ref{eq:seesaw}) by introducing a complex matrix $R$\,\cite{Casas:2001sr,Ibarra:2003up},
\begin{eqnarray}
\label{eq:solution}
 \lambda  = \frac{1}{v}\, M_R^{1/2}\, R\, {\bar m}_\n^{1/2} \, U_{MNS}^{\dagger}\ ,
\end{eqnarray}
where $R$ is given  by
\begin{eqnarray}
   R = 
\left(
\begin{array}{ccc}
0  & \cos z   & -\sin z   \\
0  & \sin z  & \cos z  
\end{array}
\right)\ ,
\end{eqnarray}
for the normal neutrino mass hierarchy and
\begin{eqnarray}
   R = 
\left(
\begin{array}{ccc}
-\sin z  & \cos z   & 0 \\
\cos z  & \sin z  & 0
\end{array}
\right)\ ,
\end{eqnarray}
for the inverted neutrino mass hierarchy.
These expressions show that a complex parameter $z$ 
accounts for the difference of the number of the parameters
in the high and the low energy theory.
In the following discussion, we parametrize the Yukawa interaction $\l$
in terms of $\bar{m}_\n$, $M_R$, $U_{MNS}$ and $z$ using Eq.\,(\ref{eq:solution}).

Before closing this section,
let us write down the elements of $\lambda$ explicitly.
For the normal neutrino mass hierarchy, the elements are given by
\begin{eqnarray}
  \l_{1\a }  &=& \frac{1} {v} \sqrt M_1 ( \sqrt{m_2}\, U^*_{\a 2}\, c_z  - \sqrt{m_3}\, U^*_{\a 3}\,s_z  ) \ ,
  \nonumber \\
   \l_{2\a  }  &=& \frac{1} {v} \sqrt M_2 ( \sqrt{m_2}\, U^*_{\a 2}\, s_z  + \sqrt{m_3}\, U^*_{\a 3}\,c_z  ) \ ,
\end{eqnarray}
and they are given by
\begin{eqnarray}
  \l_{1\a  }  &=& \frac{1} {v} \sqrt M_1 ( \sqrt{m_2}\, U^*_{\a 2}\, c_z  - \sqrt{m_1}\, U^*_{\a 1}\,s_z  ) \ ,
  \nonumber \\
   \l_{2\a  }  &=& \frac{1} {v} \sqrt M_2 ( \sqrt{m_2}\, U^*_{\a 2}\, s_z  + \sqrt{m_1}\, U^*_{\a 1}\,c_z  ) \ ,
\end{eqnarray}
for the inverted neutrino mass hierarchy $(\a = 1- 3)$.
Here, again, we have abbreviated the sine and the cosine of $z$ by $s_z = \sin z$ and $c_z = \cos z$.
It should be noted that the masses of the right-handed neutrinos
can be absorbed by $\lambda$'s by rescaling $\lambda_{i\a} \to \lambda_{i \a}/\sqrt{M_i}$.
Thus, the right-handed neutrino masses are redundant to reproduce 
the light neutrino masses and the mixings.

\section{Models with Two Texture Zeros}
\label{sec:twozeros}
To this date, the five parameters out of the seven parameters in the neutrino masses and mixings
have been measured which are summarized as\,\cite{Schwetz:2011zk},
\begin{eqnarray}
\label{eq:mass}
   {\mit \D} m_{21}^2 = 7.59^{+0.20}_{-0.18}\times 10^{-5}\, {\rm eV}^2\ , 
   &&{\mit \D}m_{31}^2 = 2.45^{+0.09}_{-0.09}\times 10^{-3}\, {\rm eV}^2\, (NH)\ , \cr
   &&  {\mit \D}m_{31}^2 = -2.34^{+0120}_{-0.09} \times 10^{-3}\, {\rm eV}^2\, (IH)\ ,
\end{eqnarray}
for the squared mass differences,
and 
\begin{eqnarray}
\label{eq:angle}
  \sin^2 \theta_{12} = 0.312^{+0.017}_{-0.015} \ ,
  &&
   \sin^2 \theta_{23} = 0.51^{+0.06}_{-0.06}\, (NH) \ ,
\,\,\,      \sin^2 \theta_{13} = 0.023^{+0.004}_{-0.004} \ , \cr
  &&
   \sin^2 \theta_{23} = 0.52^{+0.06}_{-0.06}\,(IH) \ ,
\end{eqnarray}
for the mixing angles.
The measurement of $\sin^2\theta_{13}$ is from Daya Bay\,\cite{An:2012eh},
and we do not attempt to combine the measurements of $\sin^2\theta_{13}$
at T2K\,\cite{Abe:2011sj}, MINOS\,\cite{Adamson:2011qu} and 
RENO\,\cite{Ahn:2012nd} experiments.
The errors shown above are the $1\sigma$ ranges from the 
best fit values of each parameter.
In the above lists, $NH$ denotes the normal neutrino mass hierarchy, 
and $IH$ the inverted neutrino mass hierarchy.

Now, let us try to carve the Yukawa coupling constant $\lambda$
by Occam's razor further.
For that purpose, let us first remind ourselves that 
we need at least one $CP$-phase in Eq.\,(\ref{eq:lagrangian}) 
which is required for successful leptogenesis\,\cite{Sakharov:1967dj}.
For the non-vanishing $CP$-phase, the minimal choice is given by,
\begin{eqnarray}
 \l
 =
\left(
\begin{array}{ccc}
a  & 0  & 0  \\
b  & 0   &  0 
\end{array}
\right)\ ,
\end{eqnarray}
with arbitrary exchanges of the columns and rows. 
Notice that the one phase out of two phases of the complex parameters $a$ and $b$
can be eliminated by the phase rotations of the charged leptons, $\ell_L$ and $\bar{e}_R$.
Unfortunately, however, this possibility has been excluded since it leads 
to no neutrino mixing angles and two massless neutrinos, which contradict with observations in Eq.\,(\ref{eq:angle}).
Similarly, the next minimum model for the non-vanishing $CP$-phase,
\begin{eqnarray}
 \l
 =
\left(
\begin{array}{ccc}
a  & a'  & 0  \\
b  & 0   &  0 
\end{array}
\right)\ ,
\end{eqnarray}
with an additional complex parameter $a'$ is not acceptable either, 
since it leads to two-vanishing neutrino mixing angles. 

Therefore, we need one more complex parameter in $\lambda$, or in other words,
we need a Yukawa coupling matrix with   two texture zeros.
In this case, we have four non-vanishing elements in $\lambda$
and expects one non-vanishing $CP$-phase after eliminating the three phases by 
rotating the charged leptons, $\ell_L$ and $\bar{e}_R$.%
\footnote{Here, we are assuming that $\lambda$ has no column in which 
both the elements are vanishing since it again leads to the two vanishing neutrino 
mixing angles.}
Interestingly, the Yukawa coupling $\l$ with two texture zeros for $n_N = 2$
has five free real valued parameters which correspond to the minimum necessary number 
of parameters to fit the five observed parameters in Eqs.\,(\ref{eq:mass}) and (\ref{eq:angle}).
We should emphasize that the model has only one $CP$-phase in the five parameters.
Thus, the $CP$-asymmetry required for leptogenesis and the $CP$-asymmetry
in the neutrino oscillations are related with each other\,\cite{Frampton:2002qc} (see discussions
in sec.\,\ref{sec:leptogenesis}).

Now, let us consider the condition for two texture zeros.
For the normal neutrino mass hierarchy,
the condition of $\lambda_{1\a } = 0$ is given by, 
\begin{eqnarray}
\label{eq:tanzNH}
 \tan z  = \frac{\sqrt{m_2}\, U^*_{\a 2} }{\sqrt{m_3} \, U^*_{\a 3} }\ ,
\end{eqnarray}
while the condition $\lambda_{2\a } = 0$ is given by, 
\begin{eqnarray}
 \tan z  = - \frac{\sqrt{m_3}\, U^*_{\a 3} }{\sqrt{m_2} \, U^*_{\a 2} }\ .
\end{eqnarray}
For the inverted neutrino mass hierarchy,
they are given by
\begin{eqnarray}
\label{eq:tanzIH}
 \tan z  = \frac{\sqrt{m_2}\, U^*_{\a 2} }{\sqrt{m_1} \, U^*_{\a 1} }\ ,
\end{eqnarray}
and
\begin{eqnarray}
 \tan z  = - \frac{\sqrt{m_1}\, U^*_{\a 1} }{\sqrt{m_2} \, U^*_{\a 2} }\ ,
\end{eqnarray}
respectively.
Thus, the condition for two texture zeros $\lambda_{1\a} = \lambda_{ 2\a' } = 0$
is given by
\begin{eqnarray}
\label{eq:twozeroNH}
 m_2\, U_{\a 2}\, U_{\a '2 } +  m_3\, U_{\a 3}\, U_{\a  '3 } = 0\ ,
\end{eqnarray}
for the normal neutrino mass hierarchy,
and
\begin{eqnarray}
\label{eq:twozeroIH} 
 m_2\, U_{\a 2}\, U_{\a  '2 } +  m_1\, U_{\a 1}\, U_{\a  '1 } = 0\ ,
\end{eqnarray}
for the inverted neutrino mass hierarchy.
It should be noted that the models with $\a=\a'$ again predict
two vanishing neutrino mixing angles out of the three mixing angles, 
which is inconsistent with the observations.
Thus,  in the followings, we concentrate ourselves on the models with $\a\neq \a'$.
It should be also noted that the above conditions do not depend 
on the Majorana neutrino masses, which reflects the fact
that the  Majorana neutrino masses are redundant 
for the light neutrino masses as explained in the previous section.

One may consider two texture zeros in the same row.
For the normal neutrino mass hierarchy, the condition is given by,
\begin{eqnarray}
\label{eq:twozeroNH2}
 U_{\a 2}\, U_{\a  '3 } = U_{\a 3}\, U_{\a  '2 } \ ,
\end{eqnarray}
for  $\lambda_{1\a} = \lambda_{1\a'} = 0$ 
or  $\lambda_{2\a} = \lambda_{2\a'} = 0$. 
For the inverted hierarchy,
the condition is given by,
\begin{eqnarray}
\label{eq:twozeroIH2}
 U_{\a 2}\, U_{\a '1 } = U_{\a 1}\, U_{\a '2 } \ .
\end{eqnarray}
As we will show, however, the models with two texture zeros in the same rows
cannot fit the observed masses and mixing angles consistently.

\section{Normal Hierarchy}
\label{sec:normal}
In this section, we consider the models with two texture zeros in 
the Yukawa couplings $\l$ for the normal neutrino mass hierarchy.
Here, we again emphasize that we are taking the bases where the Majorana 
neutrino masses and the charged lepton masses are diagonal.
The following analyses are the updates of the analyses in 
Refs.\,\cite{Frampton:2002qc,Ibarra:2003up,Raidal:2002xf}.
As we will show, the  seesaw mechanism with the minimal number of parameters
achieved by two texture zeros
cannot fit all the five neutrino parameters consistently
for the normal light neutrino mass hierarchy.

\subsection{Models with $\l_{1 e} = \l_{2 \m} = 0$ or $\l_{1 \m} = \l_{2 e} = 0$  $(NH)$}
From Eq.\,(\ref{eq:twozeroNH}),  the condition of two texture zeros at
$\l_{1 e}$ and  $\l_{2 \m}$ is reduced to, 
 \begin{eqnarray}
 \label{eq:condition1}
  m_3 s_{13} s_{23} e^{-i(\delta +\a )} + m_2 s_{12}
  (c_{12} c_{23} - e^{i\d} s_{12} s_{13} s_{23}) = 0 \ .
\end{eqnarray}
The condition of  two texture zeros at $\l_{2 \m}$ and  $\l_{1 e}$ is
identical to this condition.
The imaginary part of the above condition leads to a relation between
$\delta$ and $\alpha$;
\begin{eqnarray}
 \label{eq:sind}
 \sin \d = - \frac{m_3}{m_2}  \frac{1}{s_{12}^2} \sin\bar\a\ ,
\end{eqnarray}
where we have defined $\bar\a = \d+ \a$.
Thus, by remembering $s_{12}^2 \simeq 0.31$
and $m_2 \ll m_3$, we find that the Majorana phase $\bar\alpha$ $(\alpha)$ is restricted to
 \begin{eqnarray}
 \label{eq:sina21}
 |\sin\bar\a | \lesssim 0.055\ ,
\end{eqnarray}
while $\delta$ can take wide range of  values from $0$ to $2\pi$.

\begin{figure}[tb]
\begin{minipage}{.49\linewidth}
\begin{center}
  \includegraphics[width=.8\linewidth]{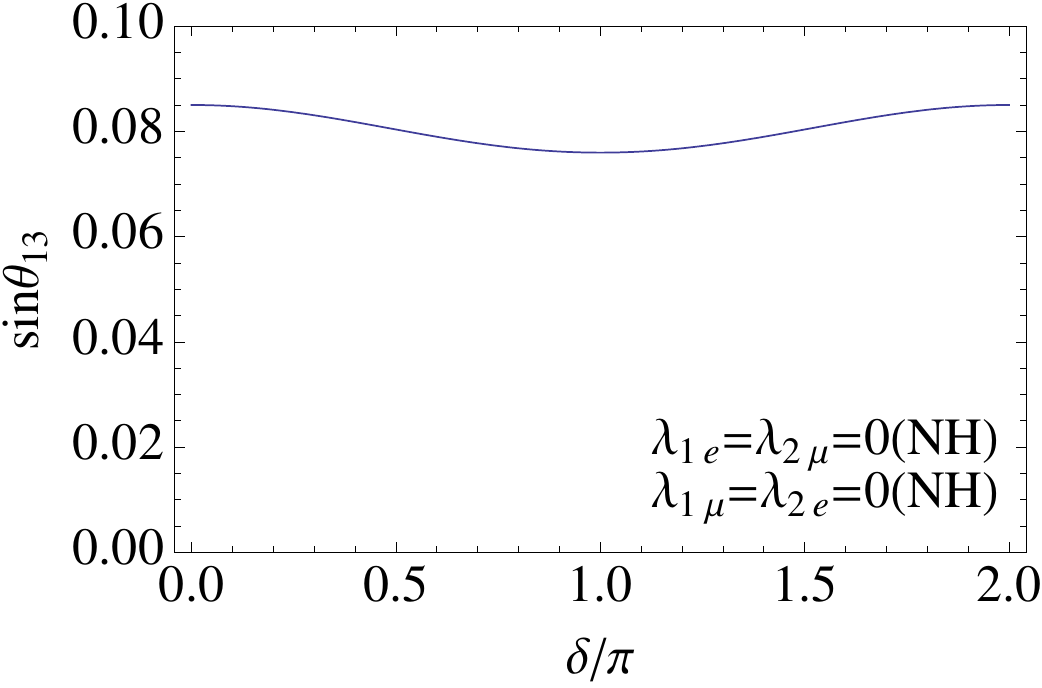}
  \end{center}
  \end{minipage}
 \begin{minipage}{.49\linewidth}
 \begin{center}
  \includegraphics[width=.8\linewidth]{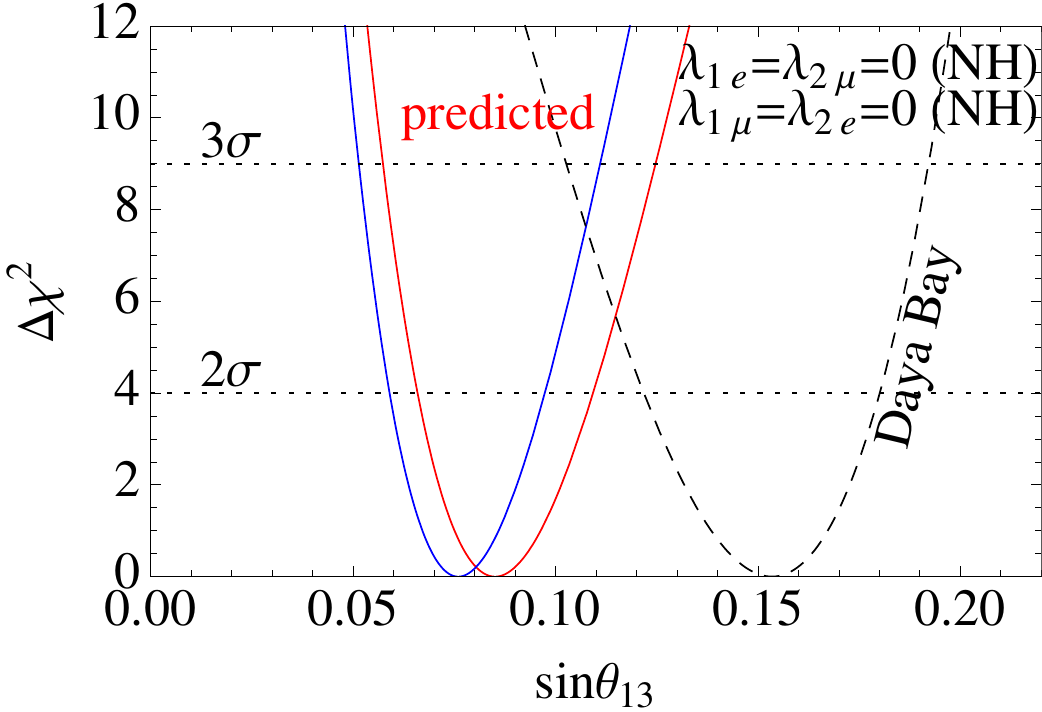}
   \end{center}
  \end{minipage}
\caption{\sl \small
(Left) The predicted range of $\sin\theta_{13}$ for the 
best fit values of the mass and mixing parameters in Eqs.\,(\ref{eq:mass})-(\ref{eq:angle})
while taking $s_{13}$ as a free parameter.
(Right) The ${\mit\Delta} \chi^2$ of the predicted $s_{13}$
for  $\d = 0$ (red) and $\d = \pi$ (blue). 
We also show ${\mit\Delta} \chi^2$ obtained at the Daya Bay experiment
as a dashed line.
}
\label{fig:21}
\end{figure}

The real part of the above condition, on the other hand, leads to,
\begin{eqnarray}
 s_{13} &=& -\frac{m_2}{m_3}\frac{c_{12}c_{23} s_{12}}{s_{23} (\cos\bar\a  - m_2/m_3\, s_{12}^2 \cos\d) } \ ,\\
             &\simeq& \frac{m_2}{m_3}\frac{c_{12}c_{23} s_{12}}{s_{23} }
             \simeq 0.08 \ .
\end{eqnarray}
where we have approximated that $(\cos\bar{\a} -  m_2/m_3\, s_{12}^2 \cos\d)  \simeq -1$ 
in view of Eq.\,(\ref{eq:sina21}) in the final expression.%
\footnote{We have defined the mixing angles from $0$ to $\pi/2$.}
Therefore, we find that the model with two texture zeros at $\l_{1e}$ and $\l_{2\m}$
(or at $\l_{1\m}$ and $\l_{2e}$) predicts $s_{13}$ 
in the range around $0.08$ for a given set of four observed parameters, 
${\mit \D}m_{21}^2$, ${\mit \D}m_{31}^2$, 
 $\sin^2\theta_{12}$, $\sin^2\theta_{23}$ in Eq.\,(\ref{eq:mass})-(\ref{eq:angle}). 
This is an unexpected result
(though already known in Refs.\,\cite{Frampton:2002qc,Ibarra:2003up,Raidal:2002xf}), 
since the models have five parameters to fit the five observed data.
This over constraint on $s_{13}$ is associated with the insensitivity 
of $s_{13}$ to $\d$ due to $m_{3}\gg m_{2}$ (see Eq.\,(\ref{eq:condition1})).
The resulting range of $s_{13}$ is, however, too small 
to be consistent with the direct measurement of $s_{13}$ in Eq.\,(\ref{eq:angle}).

In Fig.\,\ref{fig:21}, we show the range of $s_{13}$
as a function of $\delta$ for the best fit values 
of ${\mit \D}m_{21}^2$, ${\mit \D}m_{31}^2$, 
 $\sin^2\theta_{12}$, $\sin^2\theta_{23}$.
In the figure, we have solved
the real and the imaginary part of the condition  Eq.\,(\ref{eq:condition1}) for  $\bar{\alpha}$ 
to obtain $s_{13}$ as a function of $\d$.
The figure shows the insensitivity of $s_{13}$ to $\delta$.
The figure also shows that $s_{13}$ takes the maximum value 
at $\d = 0$ and the minimum value at $\d = \pi$.

In the figure, we also show ${\mit\Delta} \chi^2$ of the predicted $s_{13}$
for given values of $\delta$.
Here, we approximated that four  parameters 
${\mit \D}m_{21}^2$, ${\mit \D}m_{31}^2$, 
 $\sin^2\theta_{12}$, $\sin^2\theta_{23}$ obey
the Gaussian distribution, 
while taken $s_{13}$ as a free parameter.
By comparing ${\mit\Delta} \chi^2$ with the one obtained by the Daya Bay experiment,
we find that the prediction on $s_{13}$ is inconsistent with the observations 
at about $3\sigma$ level.
Therefore, we find that the model with two texture zeros
$\l_{1 e} = \l_{2 \m} = 0$ or $\l_{1 \m} = \l_{2 e} = 0$ 
cannot fit the five observed parameters consistently.

\subsection{Models with $\l_{1 e} = \l_{2 \t} = 0$ or $\l_{1 \t} = \l_{2 e} = 0$  $(NH)$ }
Similarly, we obtain the condition for two texture zeros,
 \begin{eqnarray}
  m_3 s_{13} c_{23} e^{-i(\delta +\a )} - m_2 s_{12}
  (c_{12} s_{23} +  s_{12} s_{13} c_{23}e^{i\d}) = 0 \ ,
\end{eqnarray}
which again leads to the same relation between $\d$ and $\bar\a$ given in Eq.\,(\ref{eq:sind}).
The real part of the above condition, on the other hand, leads to 
\begin{eqnarray}
 s_{13} &=& -\frac{m_2}{m_3}\frac{c_{12} s_{12} s_{23}}{c_{23} (\cos\bar\a  - m_2/m_3 \, s_{12}^2\cos\d) } \ ,\\
             &\simeq& \frac{m_2}{m_3}\frac{c_{12}s_{12}s_{23}}{c_{23} }
             \simeq 0.09 \ .
\end{eqnarray}
Here again, we have approximated that $(\cos\bar{\a} -  m_2/m_3 \, s_{12}^2\cos\d)  \simeq -1$ 
in view of Eq.\,(\ref{eq:sina21}).

\begin{figure}[tb]
\begin{minipage}{.49\linewidth}
\begin{center}
  \includegraphics[width=.8\linewidth]{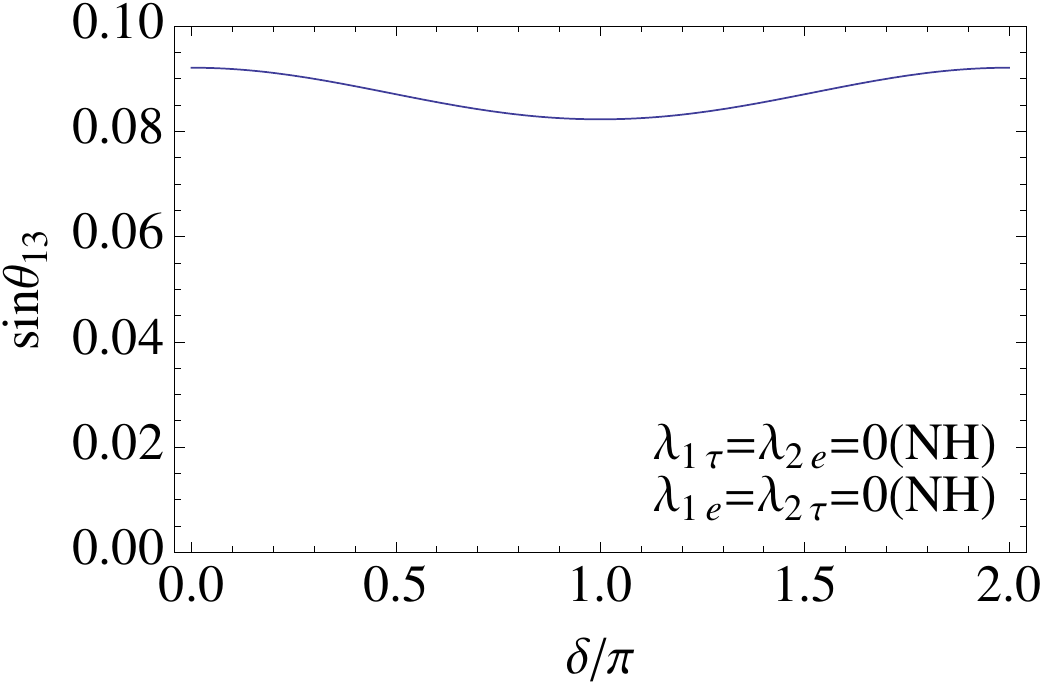}
  \end{center}
  \end{minipage}
 \begin{minipage}{.49\linewidth}
 \begin{center}
  \includegraphics[width=.8\linewidth]{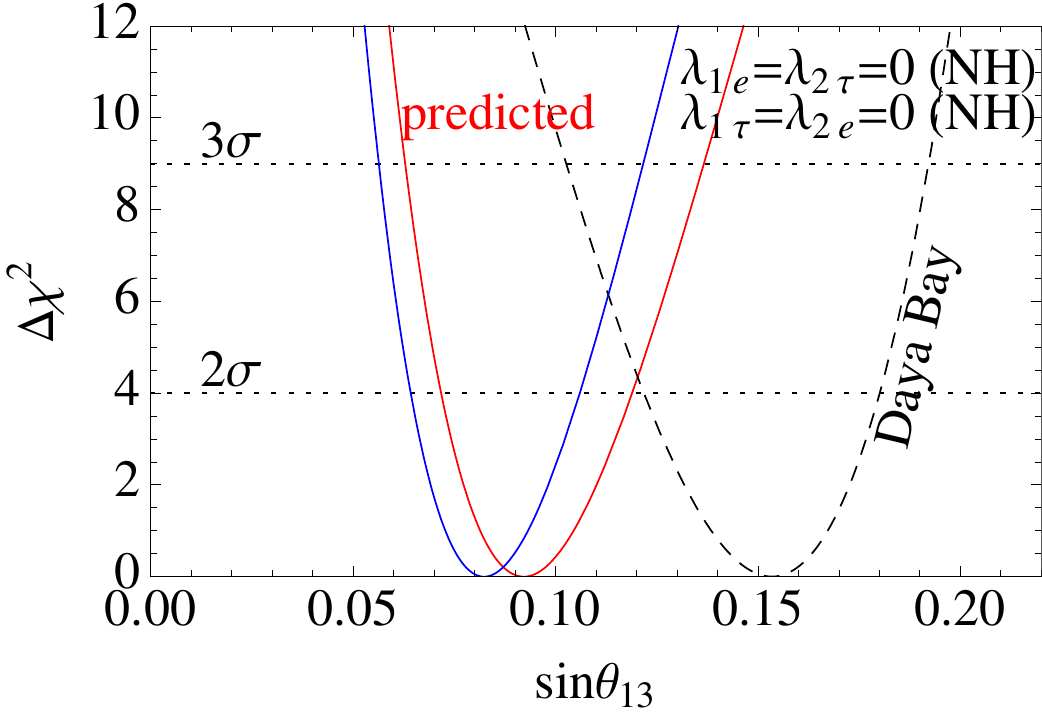}
   \end{center}
  \end{minipage}
\caption{\sl \small
(Left) The predicted range of $\sin\theta_{13}$ for the 
best fit values of the mass and mixing parameters in Eqs.\,(\ref{eq:mass})-(\ref{eq:angle})
except for the observed value of $s_{13}$. 
(Right) The ${\mit\Delta} \chi^2$ of the predicted $s_{13}$
for  $\d = 0$ (red) and $\d = \pi$ (blue). 
}
\label{fig:31}
\end{figure}

Therefore, the model predicts $s_{13}$ 
in the range around $0.09$ for a given set of four observed parameters, 
${\mit \D}m_{21}^2$, ${\mit \D}m_{31}^2$, 
$\sin^2\theta_{12}$, $\sin^2\theta_{23}$. 
In Fig.\,\ref{fig:31}, we show the predicted range of $s_{13}$ for the best fit values 
of the mass and mixing parameters in Eqs.\,(\ref{eq:mass})-(\ref{eq:angle})
except for the observed value of $s_{13}$.
We also show ${\mit\Delta} \chi^2$ of the predicted $s_{13}$
for given values of $\delta$ ($\d = 0$ or $\pi$). 
Again, we find that the predictions are inconsistent with
the observations at about the $3\sigma$ level,
by comparing ${\mit\Delta} \chi^2$ with the one obtained by Daya Bay experiment. 
Therefore, we find that the model with two texture zeros
$\l_{1 e} = \l_{2 \t} = 0$ or $\l_{1 \t} = \l_{2 e} = 0$ 
cannot fit the five observed neutrino parameters consistently.

\subsection{Models with $\l_{1 \m} = \l_{2 \t} = 0$ or $\l_{1 \t} = \l_{2 \m} = 0$  $(NH)$ }
For this texture, the condition is given by,
\begin{eqnarray}
\label{eq:condition3}
m_3 s_{23} c_{23} c_{13}^2 - m_2 (c_{12}s_{23} + c_{12}s_{13} c_{23} e^{i\d})
(c_{12}c_{23} -s_{12}s_{13} s_{23} e^{i\d}) = 0\ .
\end{eqnarray}
This equation, unfortunately, can be solved only for $s_{13}\simeq 1$ which
is inconsistent with the observed value of $s_{13}$.
To see this, it should be noted that the first term in the condition Eq.\,(\ref{eq:condition3}) is the only term 
which is proportional to $m_3$.
Thus, by remembering that $m_3 \gg m_2$ and $c_{23} \simeq s_{23} \simeq 0.5$,
we find that the condition requires $c_{13}^2  = O(m_2/m_3)$, that is $s_{13}\simeq 1$.
Therefore, models with two texture zeros  $\l_{1 \m} = \l_{2 \t} = 0$ or $\l_{1 \t} = \l_{2 \m} = 0$ 
cannot fit the neutrino parameters consistently.

\subsection{Models with $\l_{i\a}=\l_{i\a'} = 0$ ($\a\neq \a'$)  $(NH)$}
Finally, let us consider models with two texture zeros in a row.
For $\a = e$ and $\a' = \m$, the condition for two texture zeros in Eq.\,(\ref{eq:twozeroNH2}) 
is reduced to
\begin{eqnarray}
c_{12} c_{23} s_{13} - s_{12} s_{23} e^{i \d} = 0\ .
\end{eqnarray}
Thus, for $\d \neq 0$, the model predicts $s_{12} = 0$ or $s_{23} = 0$,
which is inconsistent with the observations.
For $\d = 0$, the model predicts,
\begin{eqnarray}
 s_{13} = t_{12}t_{23} \simeq 0.69\ ,
\end{eqnarray}
which is also inconsistent with the observations.

For $\a = e$ and $\a' = \t$, the condition for two texture zeros is reduced to
\begin{eqnarray}
c_{12}  s_{13} s_{23} + s_{12} c_{23} e^{i \d} = 0\ .
\end{eqnarray}
Thus, for $\d \neq 0$, the model predicts $s_{12} = 0$ or $s_{23} =10$,
which is again inconsistent with the observations.
For $\d = 0$, the model predicts,
\begin{eqnarray}
 s_{13} = t_{12}t_{23}^{-1} \simeq 0.66\ ,
\end{eqnarray}
which is also inconsistent with the observations.

For $\a = \m$ and $\a' = \t$, the condition for two texture zeros is reduced to
\begin{eqnarray}
c_{12}c_{13} = 0\ .
\end{eqnarray}
Thus, the model predicts $s_{12} = 1$ or $s_{13} =1$, which are both inconsistent
with observations.

\section{Inverted Hierarchy}
\label{sec:inverted}
In this section, we consider the models with two texture zeros 
for the inverted neutrino mass hierarchy.
Unlike the case for the normal neutrino mass hierarchy, 
we find that the model can consistently fit the current observations
in Eqs.\,(\ref{eq:mass}) and (\ref{eq:angle}) including $\sin^2\theta_{13}$.
Furthermore, we find that the models 
predict the $CP$-phase $\delta$ to be around $\delta \simeq \pi/2$ 
which can be proved/disproved in the foreseeable future.
The effective Majorana neutrino mass is also predicted around $50\,$meV 
which is also within reach of future experiments. 

\subsection{Models with $\l_{1 e} = \l_{2 \m} = 0$ or $\l_{1 \m} = \l_{2 e} = 0$ $(IH)$ }
\begin{figure}[tb]
\begin{center}
\begin{minipage}{.45\linewidth}
\begin{center}
  \includegraphics[width=.8\linewidth]{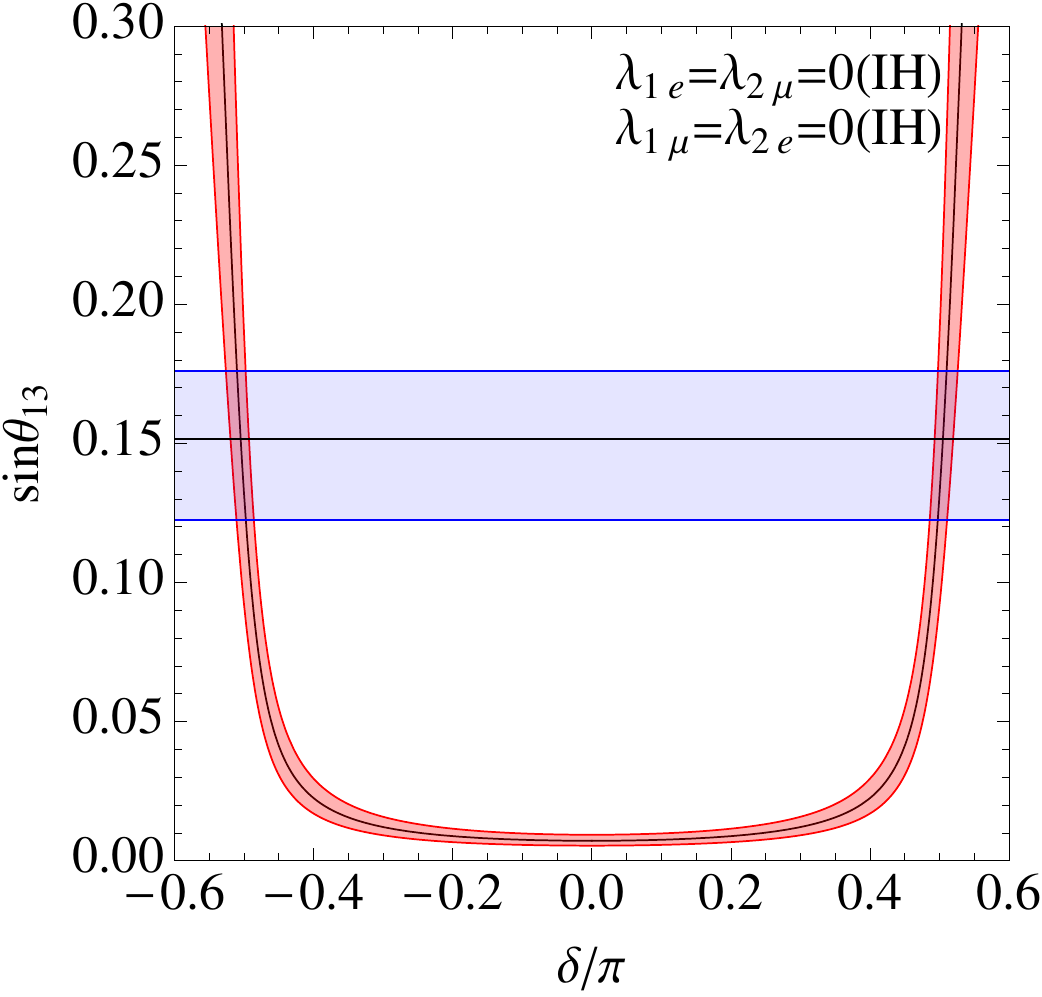}
  \end{center}
  \end{minipage}
 \begin{minipage}{.45\linewidth}
 \begin{center}
  \includegraphics[width=.8\linewidth]{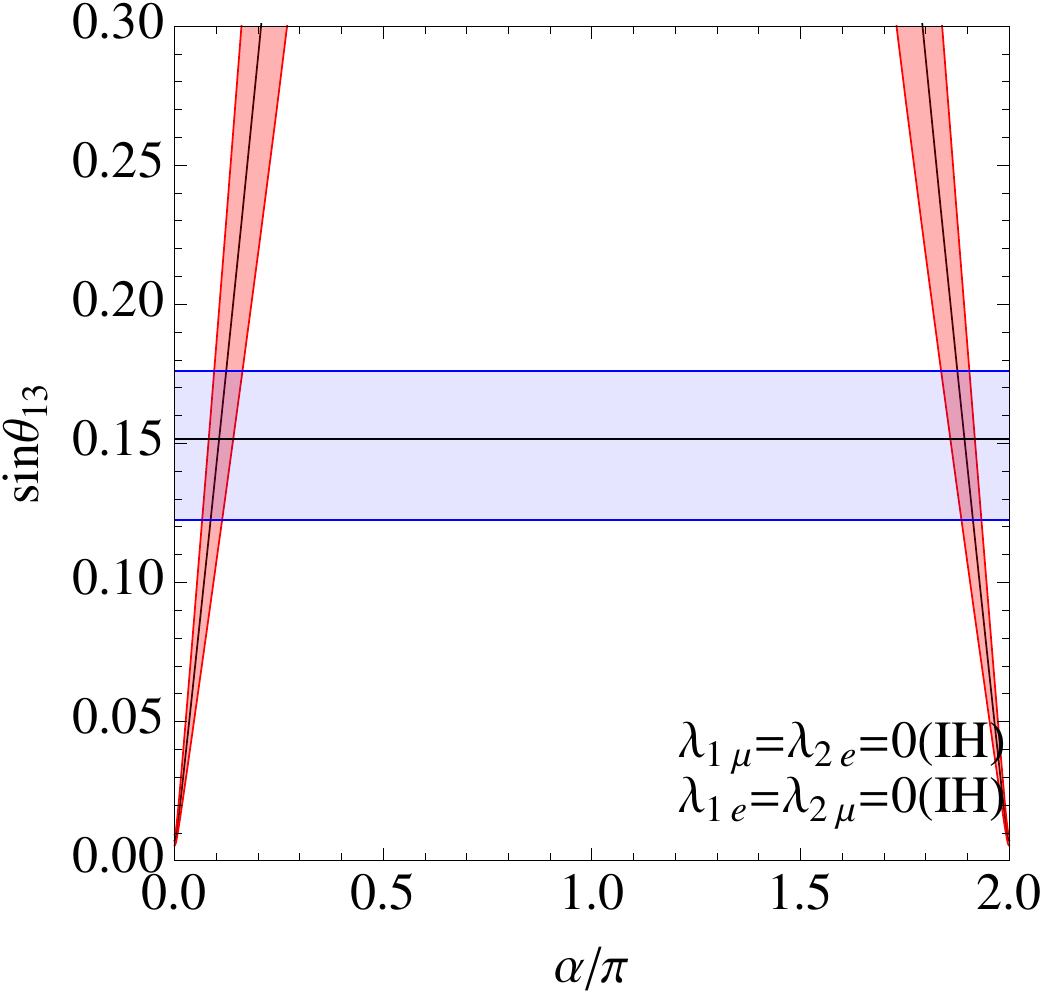}
   \end{center}
  \end{minipage}
\caption{\sl \small
(Left) The relation between $\sin\theta_{13}$ and $\delta$ 
for the inverted neutrino mass hierarchy with two texture zeros 
 $\l_{1e}=\l_{2\m}=0$ and $\l_{1\m}=\l_{2e}=0$.
(Right) The relation between $\sin\theta_{13}$ and $\alpha$. 
In both panels, we swept ${\mit \D}m_{21}^2$, ${\mit \D}m_{31}^2$, 
 $\sin^2\theta_{12}$ and $\sin^2\theta_{23}$
within the $2\sigma$ errors from the best fit values.  
The blue horizontal band shows the observed value of $\sin^2\theta_{13}$ (with $2\sigma$ errors) 
at Daya Bay.
}
\label{fig:21IH}
\end{center}
\end{figure}
For the inverted hierarchy, the condition of two texture zeros 
is given in Eq.\,(\ref{eq:twozeroIH}),
and it is given for $\l_{1 e} = \l_{2 \m} = 0$ or $\l_{1 \m} = \l_{2 e} = 0$ by,
 \begin{eqnarray}
 \label{eq:IH12}
m_1 c_{12} ( c_{23} s_{12} + c_{12} s_{23} s_{13} e^{i\d} )
- m_2 s_{12} (c_{12} c_{23} - s_{12}s_{23} s_{13} e^{i\d})e^{i\a} = 0 \ .
\end{eqnarray}
It should be noted that for the inverted hierarchy, the neutrino masses are given by 
$m_{1}=({\mit \D} m_{13}^2)^{1/2}$ and 
$m_{2} =({\mit \D} m_{13}^2+{\mit \D} m_{12}^2)^{1/2}$
which are almost degenerated with each other.
The above condition can be solved for $s_{13}$ and $\delta$ as,
\begin{eqnarray}
\label{eq:sol13}
 s_{13} e^{i\d} = - \frac{c_{12}c_{23} s_{12}(m_1 - e^{i\a} m_2 )   }
 {s_{23} (m_1 c_{12}^2  + m_2 s_{12}^2 e^{i\a})  }\ .
\end{eqnarray}
Thus, by remembering $m_1 \simeq m_2$, we find that $s_{13}$
take a wide range of values
for a given set of ${\mit \D}m_{21}^2$, ${\mit \D}m_{31}^2$, 
 $\sin^2\theta_{12}$, $\sin^2\theta_{23}$. 
In fact, the above solution can be approximated by
\begin{eqnarray}
\label{eq:sol13app}
 s_{13} e^{i\d} \simeq - \frac{c_{12}c_{23} s_{12}(1 - e^{i\a} )   }
 {s_{23}   } \simeq i  \frac{c_{12}c_{23} s_{12}   }{s_{23}}\times [\a \,\,{\rm mod}\, 2\pi ]\ ,
\end{eqnarray}
for a small $\a $ (mod $2\pi$) in the limit of $m_1 = m_2$.
From the final expression, we see that $s_{13}$ takes a wide range values,
by sweeping $\alpha$.
Therefore, all the five observed neutrino parameters 
can be consistently provided by the five parameters in the Yukawa coupling constants $\l$ 
(up to the Majorana right-handed masses).

In Fig.\,\ref{fig:21IH}, we show a predicted relation
between $\delta$ and $s_{13}$ 
for a  given set of ${\mit \D}m_{21}^2$, ${\mit \D}m_{31}^2$, 
$\sin^2\theta_{12}$ and $\sin^2\theta_{23}$.
Here, we solved Eq.\,(\ref{eq:sol13}).
In the figure, the red band corresponds 
to the $2\sigma$ errors from the best fit values  
of the above four neutrino parameters.
We also show a predicted correlation 
between $\alpha$ and $s_{13}$.
From these figures, 
we find that the model predicts the Dirac $CP$-phase 
$\delta \simeq \pm \pi/2 $ 
for $s_{13}^2 \simeq 0.023$,
while the Majorana phase $\alpha$ is rather suppressed, $\a = \pm \pi/10$ (mod $2\pi$).
This result, the large $\delta$ for a small $\alpha$, may seem strange in view of Eq.\,(\ref{eq:sol13})
where the phase of the right hand side is coming from $\alpha$.
This peculiar behavior stems from the almost degenerate two neutrino masses, i.e. $m_1 \simeq m_2$,
which leads  Eq.\,(\ref{eq:sol13app})
where the right hand side is pure imaginary even for a small $\alpha$.%
\footnote{In the exact limit of $m_1 = m_2$, the non-vanishing $\delta$ does not 
lead to any physical $CP$-violations, although we have ${\mit\D}m_{21}^2\neq 0$
in reality.}
Thus, in this model, the smallness of $\alpha$ is not related to the smallness of $\d$ 
but related to the smallness of $s_{13}$.

\begin{figure}[tb]
\begin{center}
\begin{minipage}{.45\linewidth}
\begin{center}
  \includegraphics[width=.9\linewidth]{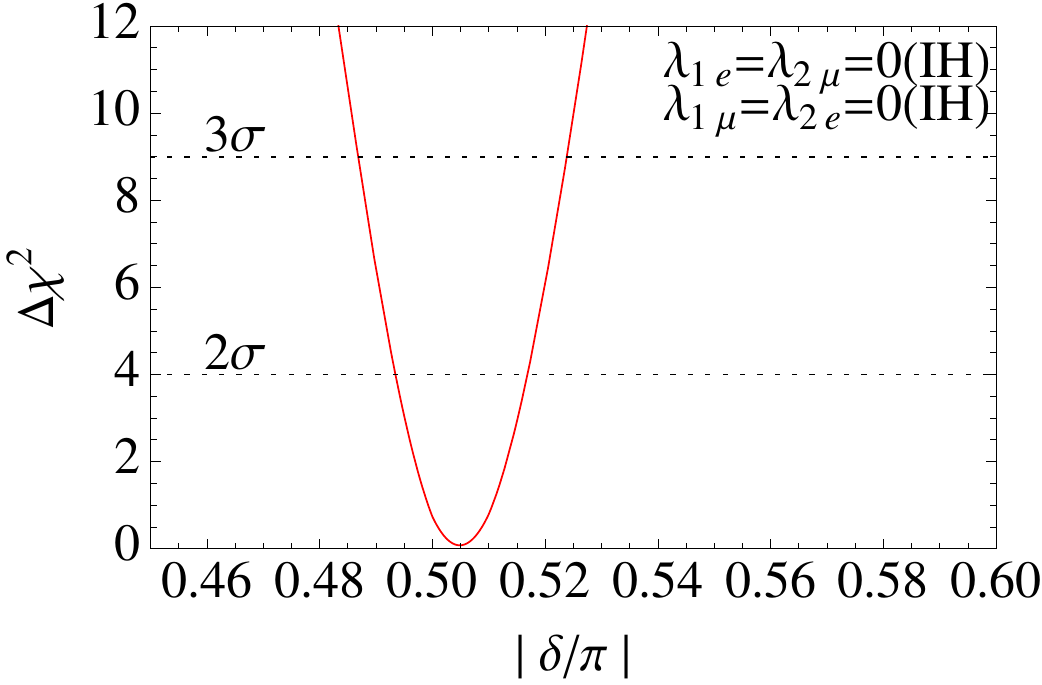}
  \end{center}
  \end{minipage}
 \begin{minipage}{.45\linewidth}
 \begin{center}
  \includegraphics[width=.9\linewidth]{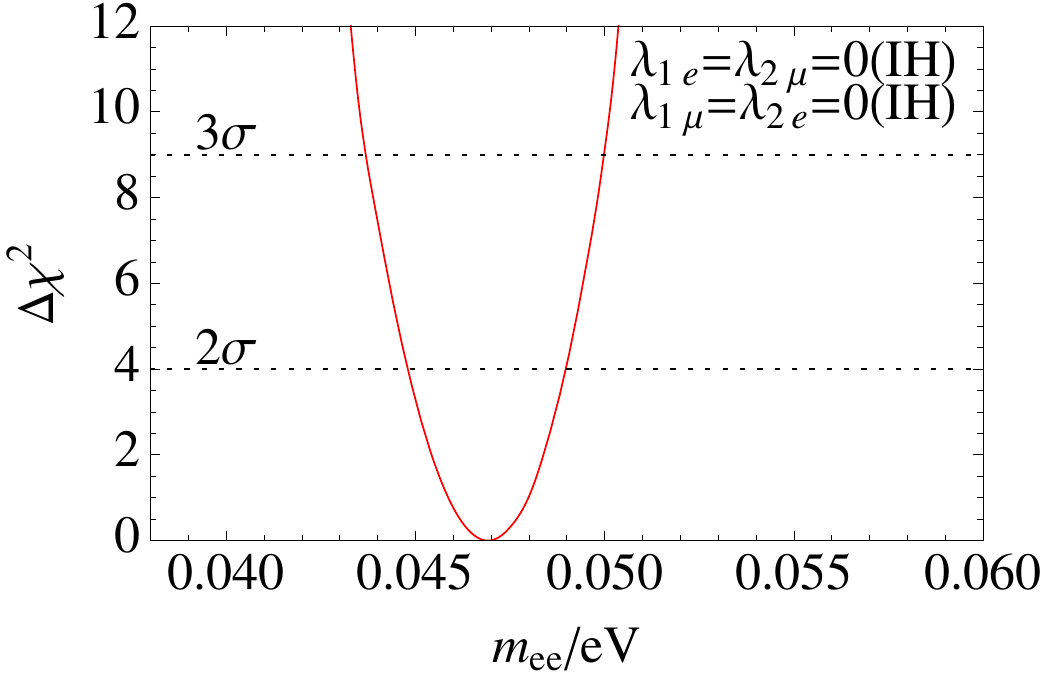}
   \end{center}
  \end{minipage}
\caption{\sl \small
(Left) The ${\mit \D}\chi^2$ of $\delta$ 
for $\l_{1e}=\l_{2\m}=0$ and $\l_{1\m}=\l_{2e}=0$.
(Right) The ${\mit \D}\chi^2$ of $m_{ee}$.
In both panels, we approximated that the five observed parameters
obey the Gaussian distributions.
}
\label{fig:21IHchisq}
\end{center}
\end{figure}

In Fig.\,\ref{fig:21IHchisq}, 
we also show ${\mit\D}\chi^2$ of the predicted $CP$-phase $\delta$
by approximating that the five observed parameters
obey the Gaussian distribution with the errors and the central values in Eq.\,(\ref{eq:mass})
and (\ref{eq:angle}).
The figure shows that the model predicts $\delta \simeq \pi/2$ very sharply.
It should be noted that such a large $CP$-phase $\delta$ can be proven/disproven
in the foreseeable future in the combination of the results of the neutrino oscillation
experiments (see for example\,\cite{Huber:2009cw}).

Since all the five model parameters relevant for the light neutrino
masses and mixing matrix have been determined 
by the observed neutrino parameters,
it is also possible to predict the rate of the neutrinoless double beta decay.
The rate of the neutrinoless double beta is proportional to the effective Majorana neutrino mass,
\begin{eqnarray}
 m_{ee} &=& | m_1 U_{e1}^2 + m_2 U_{e2}^2 + m_3 U_{e3}^2 |\ .
\end{eqnarray}
Here, $m_3 = 0$ for the inverted neutrino mass hierarchy in our case.
In Fig.\,\ref{fig:21IHchisq}, we show ${\mit\D}\chi^2$ of the predicted $m_{ee}$.
The figure shows that the model predicts a rather large value of $m_{ee}$
at around $m_{ee}\simeq \sqrt{ {\mit \D} m_{23}^2}\simeq 47$\,meV.
This value is close to the expected reaches of the coming experiments of
the neutrinoless double beta decay (see e.g. Ref.\,\cite{GomezCadenas:2010gs} and references therein).

Finally, let us explicitly write down the Yukawa coupling constants 
and the complex parameter $z$. 
For the best fit values of the observed five parameters, we find that 
the Yukawa coupling constants are given by,
\begin{eqnarray}
\lambda= \left(
\begin{array}{ccc}
0.12\times e^{-0.053\,i} & 0   & 0.028 \times e^{1.5\,i}   \\
0  &0.28\times e^{3.0\, i }   &   0.29 \times e^{-0.12\,i}
\end{array}
\right)\ ,
\end{eqnarray}
and the complex parameter $z$ is given by,
\begin{eqnarray}
 z =0.98 \times e^{-3.1\, i}\ .
\end{eqnarray}
Here, we have assumed $M_1 = 10^{13}$\,GeV and $M_2 = 10^{14}$\,GeV,
although they are redundant for fitting the low energy data.
For different Majorana masses, the Yukawa coupling constants $\l_{1\a}$
is scaled by $(M_1/10^{13}\,{\rm GeV})^{1/2}$
and $\l_{2\a}$ by $(M_2/10^{14}\,{\rm GeV})^{1/2}$.

\subsection{Models with $\l_{1 e} = \l_{2 \t} = 0$ or $\l_{1 \t} = \l_{2 e} = 0$ $(IH)$ }
Similarly, 
the condition for $\l_{1 e} = \l_{2 \t} = 0$ or $\l_{1 \t} = \l_{2 e} = 0$
is reduced to
 \begin{eqnarray}
 \label{eq:IH12}
m_1 c_{12} ( s_{12} s_{23} - c_{12}  c_{23} s_{13}e^{i\d} )
- m_2 s_{12} 
 (c_{12} s_{23} + s_{12}c_{23} s_{13} e^{i\d}) e^{i\a}= 0 \ ,
\end{eqnarray}
which can be solved for $s_{13}$ and $\delta$ as,
\begin{eqnarray}
\label{eq:sol13II}
 s_{13} e^{i\d} &=&  \frac{c_{12} s_{12}s_{23}(m_1 - e^{i\a} m_2 )   }
 {c_{23} (m_1 c_{12}^2  + m_2 s_{12}^2 e^{i\a})  }\ \,\cr
 &\simeq& \frac{c_{12} s_{12}s_{23}(1 - e^{i\a}  )   }
 {c_{23}   } 
\, \simeq \,\,-i\, \frac{c_{12} s_{12}s_{23}   }
 {c_{23}   } 
 \times [\a \,\,{\rm mod}\, 2\pi ]\ .
\end{eqnarray}
Thus again, we find that $s_{13}$ can take a wide range of values
for a given set of  ${\mit \D}m_{21}^2$, ${\mit \D}m_{31}^2$, 
 $\sin^2\theta_{12}$ and $\sin^2\theta_{23}$.
Therefore, all the five observed neutrino parameters 
can be consistently provided by the five parameters in the Yukawa coupling constants $\l$.

\begin{figure}[tb]
\begin{center}
\begin{minipage}{.45\linewidth}
\begin{center}
  \includegraphics[width=.8\linewidth]{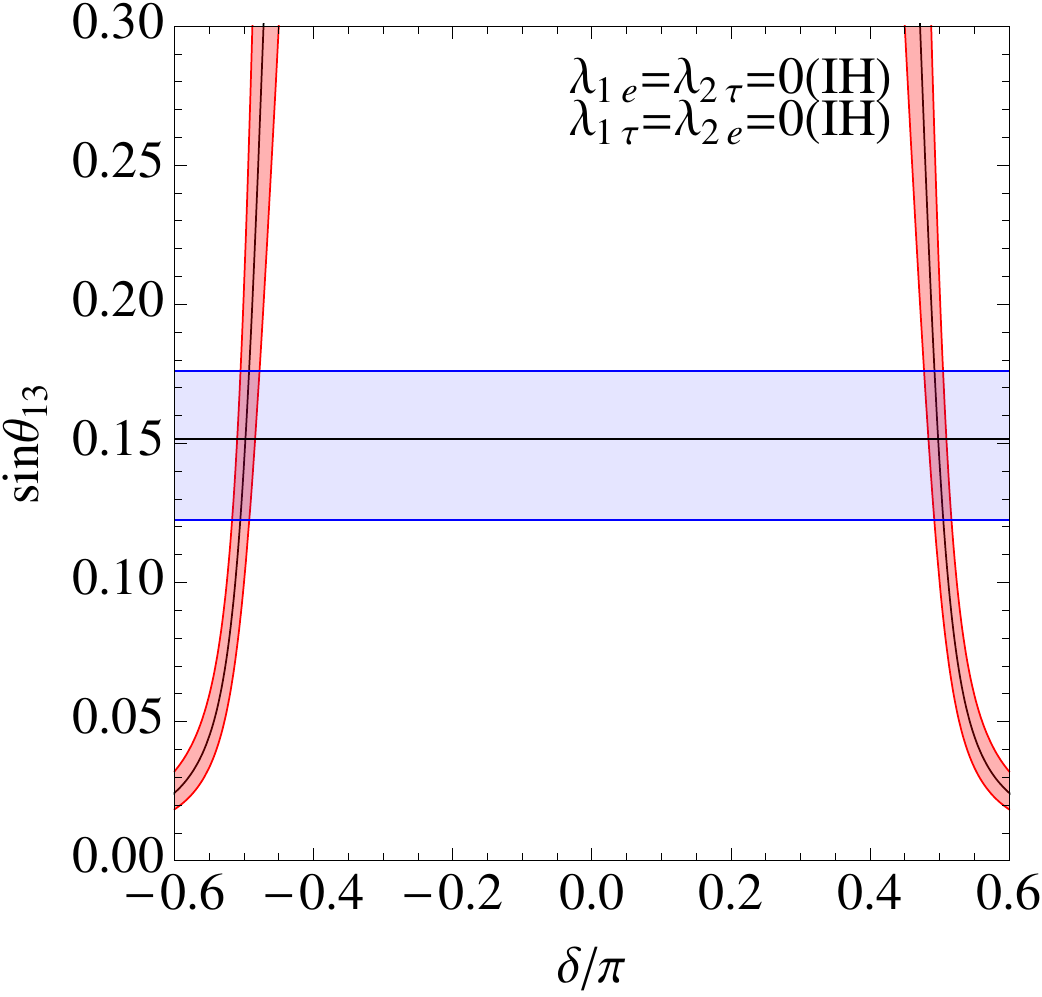}
  \end{center}
  \end{minipage}
 \begin{minipage}{.45\linewidth}
 \begin{center}
  \includegraphics[width=.8\linewidth]{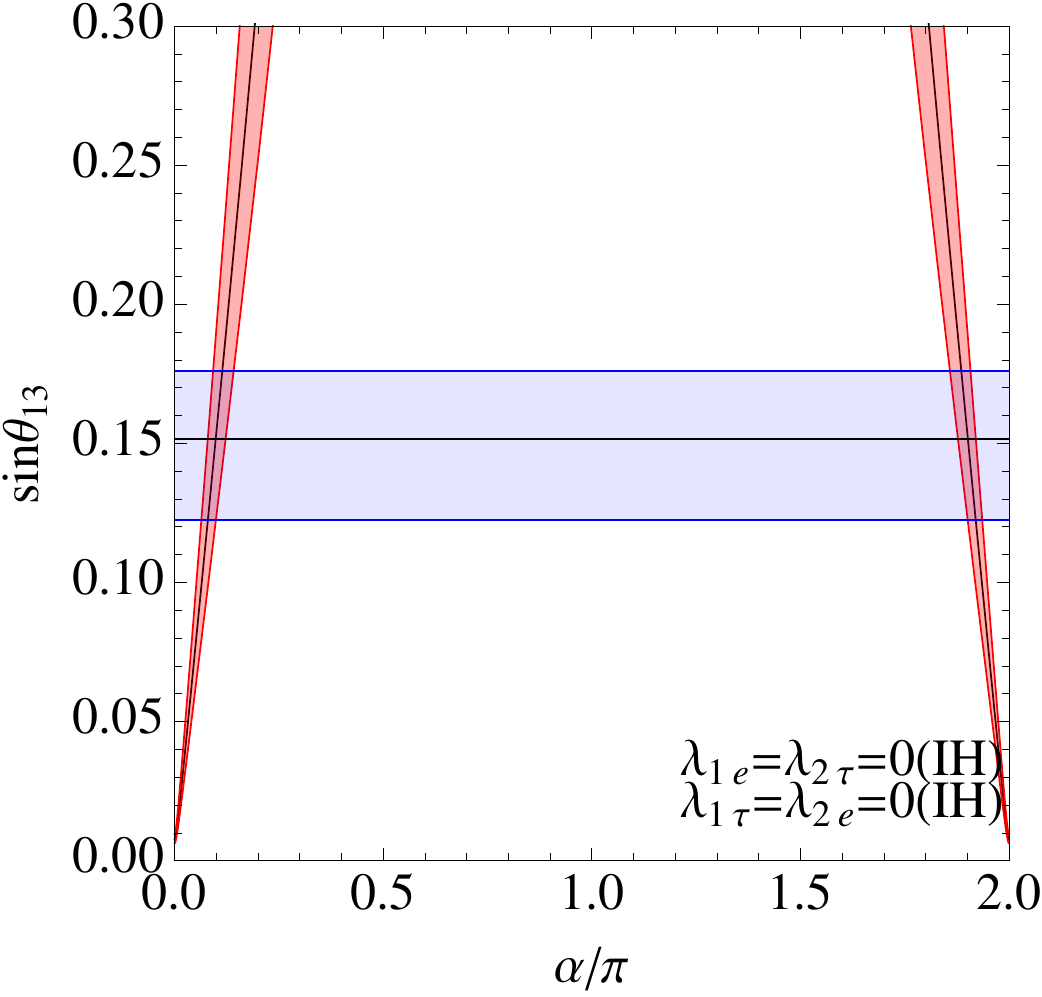}
   \end{center}
  \end{minipage}
\caption{\sl \small
(Left) The relation between $\sin\theta_{13}$ and $\delta$ 
for the inverted neutrino mass hierarchy with two texture zeros 
 $\l_{1e}=\l_{2\t}=0$ and $\l_{1\t}=\l_{2e}=0$.
(Right) The relation  between $\sin\theta_{13}$ and $\alpha$. 
See the caption of Fig.\,\ref{fig:21IH} for details.
}
\label{fig:31IH}
\end{center}
\end{figure}
\begin{figure}[tb]
\begin{center}
\begin{minipage}{.45\linewidth}
\begin{center}
  \includegraphics[width=.9\linewidth]{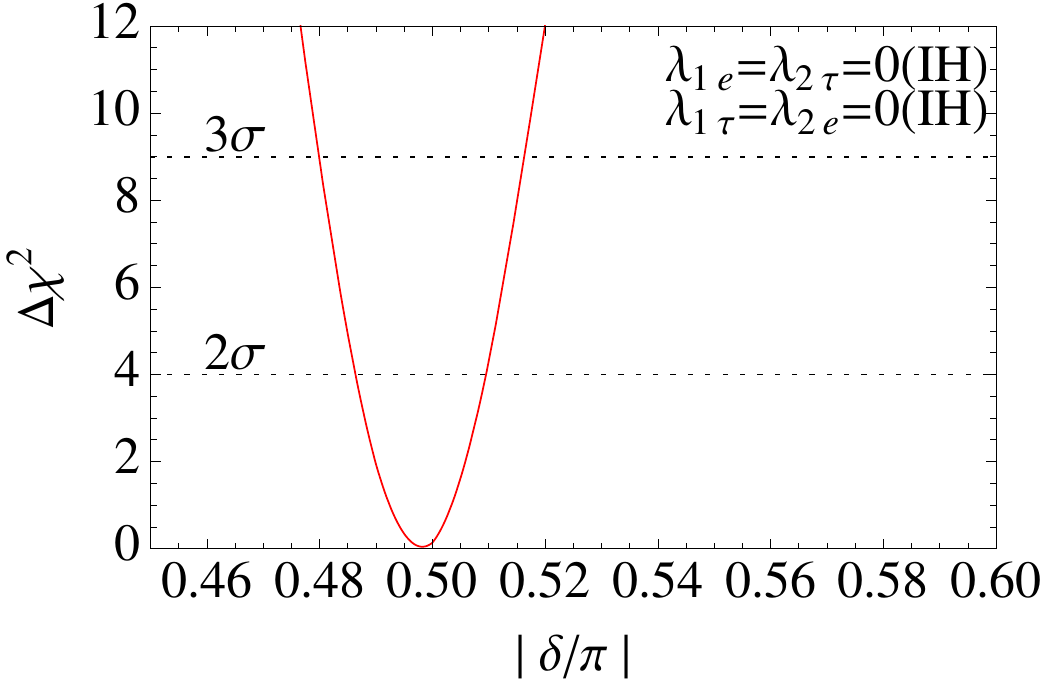}
  \end{center}
  \end{minipage}
 \begin{minipage}{.45\linewidth}
 \begin{center}
  \includegraphics[width=.9\linewidth]{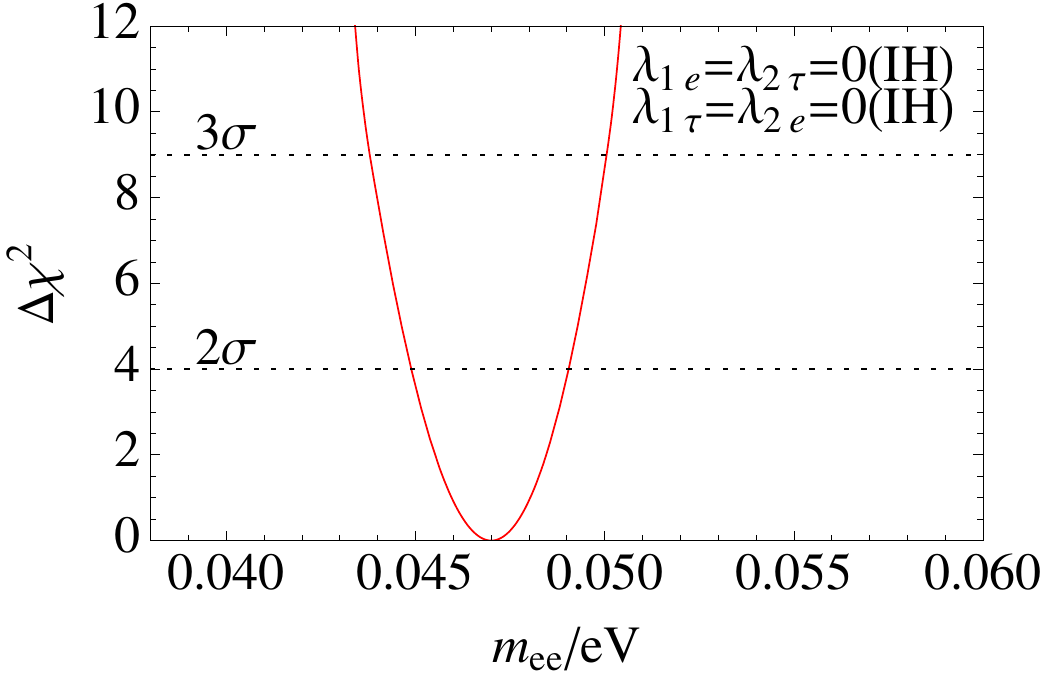}
   \end{center}
  \end{minipage}
\caption{\sl \small
(Left) The ${\mit \D}\chi^2$ of $\delta$ 
for $\l_{1e}=\l_{2\m}=0$ and $\l_{1\m}=\l_{2e}=0$.
(Right) The ${\mit \D}\chi^2$ of $m_{ee}$.
In both panels, we approximated that the five observed parameters
obey the Gaussian distributions.
}
\label{fig:31IHchisq}
\end{center}
\end{figure}

In Fig.\,\ref{fig:31IH}, 
we show a predicted relation
between $\delta$ and $s_{13}$ 
for given ${\mit \D}m_{21}^2$, ${\mit \D}m_{31}^2$, 
 $\sin^2\theta_{12}$ and $\sin^2\theta_{23}$.
We also show a predicted relation between $\alpha$ and $s_{13}$.
As a result, we again find that the model predicts the $CP$-phases 
$\delta \simeq \pm \pi/2 $  and $\alpha \simeq \pm \pi/10 $.
The smallness of $\alpha$ is again related not to 
the smallness of $\delta$ but to the smallness of $s_{13}$.
In Fig.\,\ref{fig:31IHchisq}, 
we also show ${\mit\D}\chi^2$ of the predicted $CP$-phase $\delta$.
The figure shows that the models again predict $\delta \simeq \pi/2$ very sharply,
which is within the reach of the combination of the results of the neutrino oscillation experiments.%
\footnote{It should be noted that the predicted $\delta$ here 
is slightly smaller than the one for the model with $\l_{1e} = \l_{2\m}=0$
or  $\l_{1\m} = \l_{2e}=0$.
}
In the figure, we also show  ${\mit\D}\chi^2$ of the predicted effective Majorana neutrino mass $m_{ee}$,
which is predicted around $m_{ee}\simeq 47$\,meV.

Finally, we again write down the Yukawa coupling constants 
and the parameter $z$, explicitly.
At the best fit values of the observed five parameters, we find that 
the Yukawa coupling constants are given by,
\begin{eqnarray}
\lambda= \left(
\begin{array}{ccc}
0.12\times e^{-0.049\,i} & 0.027 \times e^{-1.6\,i} & 0 \\
0  &0.28\times e^{3.0\, i }   &   0.29 \times e^{-0.11\,i}
\end{array}
\right)\ ,
\end{eqnarray}
and the complex parameter $z$ is given by,
\begin{eqnarray}
 z =0.98 \times e^{-3.1\, i}\ ,
\end{eqnarray}
which are very close to the results in the previous section.
Here, we have again assumed $M_1 = 10^{13}$\,GeV and $M_2 = 10^{14}$\,GeV,
although they are redundant for fitting the low energy data.

\subsection{Models with $\l_{1 \m} = \l_{2 \t} = 0$ or $\l_{1 \t} = \l_{2 \m} = 0$ $(IH)$ }
The condition for $\l_{1 \m} = \l_{2 \t} = 0$ or $\l_{1 \t} = \l_{2 \m} = 0$
is a bit complicated,
 \begin{eqnarray}
 \label{eq:IH12}
&&m_1 (s_{12} s_{23}-c_{12} c_{23} s_{13} e^{i \d} ) 
(c_{23} s_{12} + c_{12} s_{23}s_{13}e^{i\d}  )
 \cr
&&\,\,\,+\,
  m_2 (c_{12} s_{23}+c_{23}  s_{12} s_{13}  e^{i\d} ) 
  (c_{12} c_{23} -  s_{12} s_{23} s_{13}    e^{i \d}   )e^{i \a}
    = 0 \ .
\end{eqnarray}
Unfortunately, we have no solution to this equation
for the observed values of 
${\mit \D}m_{21}^2$, ${\mit \D}m_{31}^2$, 
 $\sin^2\theta_{12}$, $\sin^2\theta_{23}$
given in Eqs.\,(\ref{eq:mass})  and (\ref{eq:angle}).
To see this, let us rewrite the above equation into 
\begin{eqnarray}
\frac{m_2}{m_1}=-
\frac{(s_{12} s_{23}-c_{12} c_{23} s_{13} e^{i \d} ) 
(c_{23} s_{12} + c_{12} s_{23}s_{13}e^{i\d}  )}{
(c_{12} s_{23}+c_{23}  s_{12} s_{13}  e^{i\d} ) 
  (c_{12} c_{23} -  s_{12} s_{23} s_{13}    e^{i \d}   )
}e^{-i \a}\ .
\end{eqnarray}
From this expression, we find that the absolute value of the right hand side for $s_{13}\lesssim 0.3$
is approximately given by $\tan^2\theta_{12}\simeq 0.46$, which is inconsistent with 
the value of the left hand side, $m_{2}/m_{1}\gtrsim 1$.
Furthermore, we find that the absolute value of the right hand side 
takes the maximal value $1$ at $s_{13} = 1$.
Therefore, we find  no solution which satisfies the above condition.

\subsection{Models with $\l_{i\a}=\l_{i\a'} = 0$ ($\a\neq \a'$)  $(IH)$}
Finally, let us consider models with two texture zeros in a row.
For $\a = e$ and $\a' = \m$, the condition for two texture zeros in Eq.\,(\ref{eq:twozeroIH2}) 
is reduced to,
\begin{eqnarray}
 c_{13} c_{23}\, e^{i\a/2} = 0\ .
\end{eqnarray}
Thus, the model predicts $s_{13} = 1$ or $s_{23} = 1$ 
which are both inconsistent with the observations.

For $\a = e$ and $\a' = \t$, the condition for two texture zeros is reduced to
\begin{eqnarray}
c_{13} s_{23} e^{i\a/2} = 0\ ,
\end{eqnarray}
which is also inconsistent with the observations.

For $\a = \m$ and $\a' = \t$, the condition for two texture zeros is reduced to
\begin{eqnarray}
 s_{13} e^{i(\d+\a/2)} = 0\ .
\end{eqnarray}
Therefore, the model is again inconsistent with the observations.

\section{Implications on Leptogenesis}
\label{sec:leptogenesis}
In this paper, we have discussed the seesaw mechanism 
which includes the minimum number of parameters 
for successful leptogenesis and three neutrino oscillations 
in the spirit of Occam's razor.
As a result, we found that the models with two texture zeros 
can successfully fit all the five observed neutrino parameters
by using five model parameters for the inverted neutrino mass hierarchy.
Furthermore, we also found that the model predicts the maximal $CP$-phases 
in the neutrino mixing matrix.
This result is very encouraging for leptogenesis 
which requires non-vanishing $CP$-phase.
In this section, we discuss numerical implications of the models with two texture
zeros on leptogenesis.

We assume the first right-handed neutrino $N_1$ is much lighter than
$N_2$ for simplicity. 
The baryon asymmetry generated by leptogenesis is approximately given by\,\cite{Giudice:2003jh},
\begin{eqnarray}
 \eta_{B_0} = n_B/n_\gamma \simeq   -3.4\times10^{-4}\times \varepsilon_1 
 \left( \frac{0.01\,{\rm eV}}{\tilde m_1}\right)^{1.16}\ ,
\end{eqnarray}
which fits well the numerical result for $\tilde m_1 \gtrsim 10^{-2}\,$eV.%
\footnote{
For the degenerated right-handed neutrino masses,
the baryon asymmetry can be enhanced (for recent developments see \cite{Garbrecht:2011aw,Garny:2011hg}).
}
Here, $\tilde{m}_1$ is the so-called effective neutrino mass which is 
related to the decay rate of the lighter right-handed neutrino, 
\begin{eqnarray}
\label{eq:mtilde}
 \tilde{m}_1  = \sum_\ell |\lambda_{1\ell}|^2\frac{v^2}{M_1}\ ,
\end{eqnarray}
and $\varepsilon_L$ is the $CP$-asymmetry at the decay of the lighter right-handed neutrino,
 \begin{eqnarray}
 \varepsilon_1 = -\frac{3}{16\pi} \frac{M_1}{(\l \l^\dagger)_{11}} 
{\rm Im} [( \lambda \lambda^\dagger M_R^{-1} \lambda^* \lambda^T)_{11}]\ .
\end{eqnarray}

In terms of the low energy parameter and the complex parameter $z$, the above two parameters 
for leptogenesis can be rewritten as follows. 
The coefficient for the tree-level decay width can be reduced to
\begin{eqnarray}
\lambda \lambda^\dagger &=&
\frac{1} {v^2} M_R^{1/2} R \bar{m}_\n R^\dagger M_R^{1/2}\ , \cr
&=&
\frac{1}{v^2}
\left(
\begin{array}{cc}
M_1 (m_1 |s_z|^2 + m_2 |c_z|^2)  &    \sqrt{M_1 M_2}(-m_1 c_z^* s_z + m_2 s_z^* c_z)  
\\
 \sqrt{M_1 M_2}(-m_1 c_z s_z^* + m_2 s_z c_z^*)  
   &   M_2 (m_1 |c_z|^2 + m_2 |s_z|^2) 
\end{array}
\right)\ .
\end{eqnarray}
The numerator of the $CP$-asymmetry is also reduced to
\begin{eqnarray}
\lambda\lambda^\dagger M_R^{-1} \lambda^* \lambda^T
&=& \frac{1}{v^4}
M_R^{1/2} R\, \bar{m}_\n^2  R^T M_R^{1/2} \ , \cr
&=&
\frac{1}{v^4}
\left(
\begin{array}{cc}
M_1 (m_1^2 s_z^2 + m_2^2 c_z^2)  &    \sqrt{M_1 M_2}(-m_1^2 + m_2^2 )  c_z s_z
\\
 \sqrt{M_1 M_2}(-m_1^2 + m_2^2 ) c_z s_z
   &  M_2 (m_1^2 c_z^2 + m_2^2 s_z^2) 
\end{array}
\right)\ .
\end{eqnarray}
Therefore, the parameters $\tilde m_1$ and  $\varepsilon_1$ can be expressed by,
\begin{eqnarray}
\tilde m_1 = (m_1 |s_z|^2 + m_2 |c_z|^2)\ ,
\end{eqnarray}
and
\begin{eqnarray}
\varepsilon_1 = -\frac{3}{16\pi}\frac{M_1}{v}\frac{{\rm Im}[m_1^2 s_z^2  + m_2^2 c_z^2 ]}
{v(m_1 |s_z|^2 + m_2 |c_z|^2)} \ .
\end{eqnarray}
It should be noted that $\varepsilon_1$ is proportional to ${\mit \D} m_{12}^2$,
since  ${\rm Im}[m_1^2s_z^2 + m_2^2 c_z^2 ] = {\mit \D}m_{12}^2\times {\rm Im}[c_z^2]$,
and hence, the $CP$-asymmetry is rather suppressed.

From these expressions, we find that 
the effective neutrino mass is given by
\begin{eqnarray}
 \tilde{m}_1 = (4.9 \pm 0.1)\times 10^{-2}\,{\rm eV}\ ,
\end{eqnarray}
and the $CP$-asymmetry is given by 
\begin{eqnarray}
\varepsilon_1 \simeq \pm(2.2 \pm 0.3) \times10^{-6} \left(\frac{M_1}{10^{13}\,\rm GeV}\right)\ ,
\end{eqnarray}
for sign$(\delta)=\mp 1$ for model with $\lambda_{1e}=\lambda_{2\m} = 0$.
Here, the errors correspond to $1\sigma$ errors in Eqs.\,(\ref{eq:mass}) and (\ref{eq:angle}).
As a result, the baryon asymmetry is given by,
\begin{eqnarray}
 \eta_{B_0} \simeq \pm\, 5.9 \times 10^{-10} \times \left(\frac{M_1}{5\times 10^{13}\, \rm GeV}\right)\ ,
\end{eqnarray}
for sign$(\delta)=\pm 1$.
Therefore, we find that the observed baryon asymmetry $\eta_{B_0} = (6.19\pm 0.15)\times 10^{-10}$\,\cite{Komatsu:2010fb} can be successfully generated by leptogenesis for sign$(\delta)=+1$
and $M_{1}=O(10^{13})$\,GeV.%
\footnote{
In \cite{petcov}, it has been pointed out that
there is a lower bound on $s_{13 }\sin\d$, $|s_{13}\sin\d| \gtrsim 0.11$,
for successful leptogenesis for a small Majorana $CP$-phase, $\a$.
In our model, this condition is satisfied due to 
the maximal Dirac $CP$-phase, $|\sin\d| \simeq 1$.
}
It should be noted that the correlation between the signs of $\delta$ and $\epsilon_1$
is opposite for the model with $\lambda_{1\m}=\lambda_{2e} = 0$.
Thus, unfortunately, it is not possible to predict the sing of $\delta$ from 
the known sign of the baryon asymmetry via leptogenesis.
 
Similarly, for the model with $\lambda_{1e}= \lambda_{2\tau} = 0$, 
the effective neutrino mass and  the $CP$-asymmetry are given by,
\begin{eqnarray}
 \tilde{m}_1 = (4.9 \pm 0.1)\times 10^{-2}\,{\rm eV}\ ,
\end{eqnarray}
and
\begin{eqnarray}
\varepsilon_1 \simeq \pm(2.0 \pm 0.3) \times10^{-6} \left(\frac{M_1}{10^{13}\,\rm GeV}\right)\ ,
\end{eqnarray}
for sign$(\delta)=\pm 1$.
Here again, the errors correspond to $1\sigma$ errors in Eqs.\,(\ref{eq:mass}) and (\ref{eq:angle}).
Thus, again, we find that the predicted baryon asymmetry is given by,
\begin{eqnarray}
 \eta_{B_0} \simeq \mp\, 6.5 \times 10^{-10} \times \left(\frac{M_1}{6\times 10^{13}\, \rm GeV}\right)\ ,
\end{eqnarray}
for sign$(\delta)=\pm 1$.
Therefore, the observed baryon asymmetry can be explained by leptogenesis sign$(\delta)=-1$
and $M_{1}\simeq 10^{13}$\,GeV.
The correlation between the signs of $\delta$ and $\eta_{B_0}$ is again opposite 
for $\l_{1\t} = \l_{2e} = 0$.

Finally, let us emphasize the relation between the $CP$-asymmetry for leptogenesis
and the $CP$-violations in the neutrino oscillations\,\cite{Frampton:2002qc}.
As we have seen, the $CP$-asymmetry for leptogenesis is proportional to Im\,$[ c_z^2]$,
where $\tan z$ is given by using Eqs.\,(\ref{eq:tanzIH}) and (\ref{eq:sol13}), 
\begin{eqnarray}
\label{eq:tanz}
\tan z &=& e^{-i\frac{\a}{2}}\, \tan\theta_{12}
\times \left(\frac{m_2}{m_1}\right)^{1/2} \ , \cr
\tan z &=& -  e^{i\frac{\a}{2}}\, \cot\theta_{12}
\times \left(\frac{m_1}{m_2}\right)^{1/2} \ ,
\end{eqnarray}
for the models with $\lambda_{1e}=\lambda_{2\m} = 0$ or $\lambda_{1\m}=\lambda_{2e} = 0$,
respectively.
The same expressions of $\tan z$ are also obtained
for the model with $\lambda_{1e}=\lambda_{2\t} = 0$ or $\lambda_{1\t}=\lambda_{2e} = 0$
by using  Eqs.\,(\ref{eq:tanzIH}) and (\ref{eq:sol13II}).
The $CP$-violations in the neutrino oscillations are, on the other hand, given by the Dirac $CP$-phase,
$\delta$, which can be rewritten as a Jarlskog invariant\,\cite{Jarlskog:1985cw},
\begin{eqnarray}
 J_{CP} = {\rm Im}\,[U_{\m 3} U_{e3}^* U_{e2} U_{\m 2}^*] = 
c_{12} c_{23}  c_{13}^2 s_{12} s_{23} s_{13} \sin\d \simeq 0.034 \times \sin\d \ .
\end{eqnarray}
These two $CP$-violations are related by Eqs.\,(\ref{eq:sol13}) and (\ref{eq:tanz}), which leads to,
\begin{eqnarray}
\label{eq:cz1}
{\rm Im}[ c_z^2]  = \pm s_{12} c_{12} t_{23} s_{13} \sin\delta = \pm\frac{J_{CP}}{c_{13}^2 c_{23}^2}\ ,
\end{eqnarray}
for $\l_{1e}= \l_{2\m} = 0$ (plus) and for $\l_{1\m}= \l_{2e} = 0$ (minus), respectively.
Similarly, we obtain
\begin{eqnarray}
\label{eq:cz2}
{\rm Im}[ c_z^2]  = \mp s_{12} c_{12} t_{23}^{-1} s_{13} \sin\delta =\mp \frac{J_{CP}}{c_{13}^2 s_{23}^2}\ ,
\end{eqnarray}
for $\l_{1e}= \l_{2\t} = 0$ (minus) and for $\l_{1\t}= \l_{2e} = 0$ (plus), respectively.
Therefore, we find that the $CP$-asymmetry for leptogenesis 
is interrelated to the $CP$-violations in the neutrino oscillations.
In other words, we may even say that the observed baryon asymmetry 
predicts the existence  of the $CP$-violations in the neutrino oscillations.
This reflects the fact that the model has only one $CP$-phase in
the models with two right-handed neutrinos and two texture zeros.

We also emphasize that the non-vanishing $CP$-asymmetry for
leptogenesis is guaranteed by the observed value of $s_{13}$, i.e. $s_{13}^2 = 0.023\pm 0.004$.
As we see from Eqs.\,(\ref{eq:cz1}) and (\ref{eq:cz2}), the $CP$-asymmetry 
for leptogenesis is vanishing either for $s_{13} = 0$ or $\sin\d = 0$.
Interestingly, however,  
the observed value of $s_{13}$,  $s_{13}^2 = 0.023\pm 0.004$,
predicts $\sin\d \simeq \pm 1$ (see the right panels of Fig.\,\ref{fig:21IH} and Fig.\,\ref{fig:31IH}).
Therefore, we find that the non-vanishing $CP$-asymmetry, 
$\epsilon_1 \propto {\mit \D}m_{21}^2\times {\rm Im}\,[c_z^2]$,
is guaranteed by the observed value of $s_{13}$.%
\footnote{
In other words, the predicted $s_{13}$ is too small,
$s_{13}\simeq (7-8)\times 10^{-3}$, for  a vanishing $CP$-phase in the Yukawa coupling.
Namely, the observed $\theta_{13}$ already shows non-vanishing $CP$-violation at
the  high energies.
}

\section{Conclusions and Discussions}
In this paper, we have discussed the seesaw mechanism 
which includes the minimum number of parameters 
for successful leptogenesis and the three neutrino oscillations,
in the spirit of Occam's razor.
As a result, we found that the  seesaw mechanism with the minimal number of parameters
achieved by two texture zeros can fit  all the five observed neutrino parameters consistently
for the inverted light neutrino mass hierarchy, while 
it is not possible for the normal neutrino mass hierarchy.
As interesting  predictions, we found that the model predicts the maximal Dirac 
$CP$-phase, $\d \simeq \pm \pi/2$ 
in the neutrino mixing matrix which can be measurable in the foreseeable future for the inverted neutrino mass hierarchy.
The model also predicts the effective Majorana mass, $m_{ee}\simeq 50$\,meV, responsible
for the neutrinoless double beta decay close to the reaches 
of coming experiments.

We also comment that the predicted total mass of the neutrinos 
can be probed by cosmological and astrophysical measurements.
As we have seen, the model can be consistent with the observations
only for the inverted neutrino mass hierarchy which says,
\begin{eqnarray}
 m_1 = 4.84^{+0.10}_{-0.09}\times 10^{-2}\,{\rm eV}\ , \quad
  m_2 = 4.92^{+0.10}_{-0.09}\times 10^{-2}\,{\rm eV}\ , 
\quad m_3 = 0\ ,
\end{eqnarray}
with $1\sigma$ errors.
Here, the vanishing third neutrino mass comes from the fact that 
only two right-handed neutrinos contribute to the seesaw mechanism
in our case.
Therefore, the total neutrino mass is predicted to be (at the $1\sigma$ level)
\begin{eqnarray}
\sum_{i=1-3} m_{i} = 9.75_{-0.19}^{+0.20}\times 10^{-2}\,{\rm eV}\ .
\end{eqnarray}
This total mass is within the reach of the future cosmological 
and astrophysical measurements, such as the CMB spectrum,
the galaxy distributions,  and the redshifted  $21$cm line 
in the foreseeable future (see \cite{Abazajian:2011dt} and references therein).

Several other comments are in order.
In our arguments, we have not considered any origins 
of the peculiar structure of two texture zeros. 
Obviously, it is not easy to explain such texture by simple symmetries.
Therefore, it will be puzzling if the predicted Dirac $CP$-phase
and the effective Majorana mass predicted in this model 
are confirmed in the future experiments.

\begin{figure}[tb]
\begin{center}
\begin{center}
  \includegraphics[width=.3\linewidth]{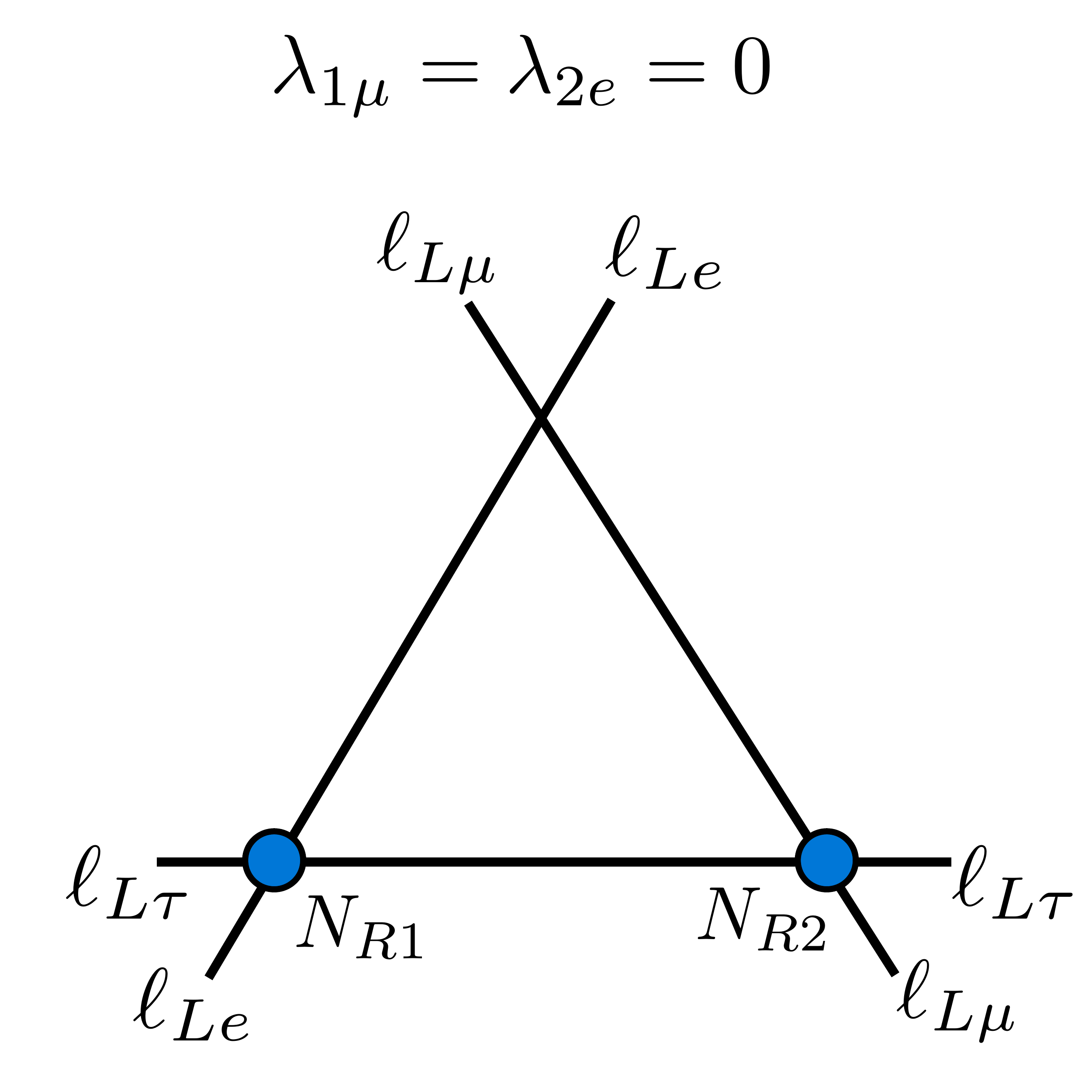}
  \end{center}
\caption{\sl \small
An illustrative picture of the extra dimensional realization
of two texture zeros.
Here, we are assuming that charged leptons $\ell_L$ and $\bar{e}_R$ are confined 
on the branes in the extra dimensions shown as lines in the figure, 
while two right-handed neutrinos $N_{R1,2}$
reside on the intersections of the branes.
We are also assuming that the Higgs boson is not localized on the branes.
}
\label{fig:extra}
\end{center}
\end{figure}

As a possible origin of two texture zeros, one may consider 
extra-dimensional models.
That is, by assuming separations of the leptons and/or Higgs bosons in the extra dimensions,
models can have texture zeros.
In Fig.\,\ref{fig:extra}, we show an illustrative picture of such separations
in the extra dimensional model.
In this illustrative example, two texture zeros are realized 
while the right-handed Majorana mass and the charged lepton
masses are diagonal
(see \cite{Raidal:2002xf} and references therein for other realizations 
of two texture zeros).

Besides the puzzle on the origins of texture zeros,
one may concern about the third right-handed neutrino
whose existence is naively expected in view of the three generations
of the quark sector and charged lepton sector.
One simple realization of the models with two right-handed neutrinos
out of three right-handed neutrinos is to assume very heavy third right-handed neutrino. 
By remembering that the contributions of the right-handed neutrinos to the neutrino mass 
in Eq.\,(\ref{eq:seesaw}) are proportional to the inverse of their masses, 
the third right-handed neutrino contributions are vanishing in its heavy mass limit, 
which effectively leads to the models with two right-handed neutrinos.

Another but more ambitious possibility is the other limit, i.e. the very light third right-handed 
neutrino with the mass in ones to tens keV range\,\cite{Kusenko:2010ik}.
In this case, if the third right-handed neutrino couples to the other fields very weakly,
it may have a lifetime much longer than the age of universe. 
Such a ``sterile neutrino" is a viable candidate of dark matter\,\cite{Dodelson:1993je}.%
\footnote{For models of sterile neutrino dark matter where all the three generations
of the right-handed neutrinos are below the electroweak scale, see 
Refs.\,\cite{deGouvea:2005er,Asaka:2005an}. }
Especially, if the third right-handed neutrino is dark matter, the constraints from X-ray observations
have put sever upper limits on the effective mixing angle $\theta$ between
the light neutrinos and the third right-handed neutrino%
\footnote{Here, the square of the effective mixing angle is 
defined by $\h^2 = \tilde{m}_3/M_3$ where $\tilde{m}_3$
is given by Eq.\,(\ref{eq:mtilde}) with the replaced indices from $1$ to $3$.
} 
such that $\theta^2 < O(10^{-10})$ for $M_{3} \simeq 10$\,keV for example (see 
e.g. \cite{Herder:2009im} for details and \cite{Boyarsky:2009ix,Kusenko:2009up} for reviews).
For this range of the small mixing, the contributions of the third right-handed 
neutrino to $m_\n$ in Eq.\,(\ref{eq:seesaw}) is negligible, and hence,
the model effectively consists of two-right handed neutrinos for the seesaw mechanism.

\section*{Acknowledgments}
We appreciate Serguey Petcov for useful comments on leptogenesis.
This work is supported by Grant-in-Aid for Scientific research from the Ministry of Education, Science, Sports, and Culture (MEXT), Japan, No.\ 22244021 (T.T.Y.),
No.\ 24740151 (M.I),  and also by World Premier International Research Center Initiative (WPI Initiative), MEXT, Japan.
 The work of K.H. is supported in part by JSPS Research Fellowships for Young Scientists.

\end{document}